\documentclass[english]{article}
\usepackage[T1]{fontenc}
\usepackage[latin9]{inputenc}
\usepackage{color}
\usepackage{array}
\usepackage{units}
\usepackage{multirow}
\usepackage{amsthm}
\usepackage{amsmath}
\usepackage{amssymb}
\usepackage{wasysym}
\usepackage{graphicx}
\usepackage{esint}

\makeatletter

\providecommand{\tabularnewline}{\\}

\numberwithin{equation}{section}
\numberwithin{figure}{section}

\usepackage{babel}
\usepackage{fullpage}
\usepackage{authblk}
\usepackage{hyperref}
\usepackage{placeins}
\usepackage{hyphenat}
\graphicspath{./figure/}
\usepackage{fancyvrb}

\title{Pricing and Hedging GMWB in the Heston and in the Black-Scholes
with Stochastic Interest Rate Models}
\author{ \textsc{Ludovic Goudenege}\thanks{F\'ederation de Math\'ematiques de l'\'Ecole Centrale Paris - CNRS FR3487 - \texttt{ ludovic.goudenege@math.cnrs.fr}} \and \textsc{Andrea Molent}\thanks{DEAMS, Universit\`a di Trieste - \texttt{andrea.molent@phd.units.it}} \and \textsc{Antonino Zanette}\thanks{Dipartimento di Scienze Economiche e Statistiche, Universit\`a di Udine - \texttt{antonino.zanette@uniud.it}}}
\date{}
\definecolor{blue}{rgb}{0,0.2,0.55}

\makeatother

\usepackage{babel}
\begin{document}
\maketitle

\begin{center}
\textbf{Abstract}
\par\end{center}

Valuing Guaranteed Minimum Withdrawal Benefit (GMWB) has attracted
significant attention from both the academic field and real world
financial markets. As remarked by Yang and Dai \cite{YD}, the Black
and Scholes framework seems to be inappropriate for such a long maturity
products. Also Chen Vetzal and Forsyth in \cite{CVF} showed that
the price of these products is very sensitive to interest rate and
volatility parameters. We propose here to use a stochastic volatility
model (Heston model) and a Black Scholes model with stochastic interest
rate (Hull White model). For this purpose we present four numerical
methods for pricing GMWB variables annuities: a hybrid tree-finite
difference method and a Hybrid Monte Carlo method, an ADI finite difference
scheme, and a Standard Monte Carlo method. These methods are used
to determine the no-arbitrage fee for the most popular versions of
the GMWB contract, and to calculate the Greeks used in hedging. Both
constant withdrawal, optimal surrender and optimal withdrawal strategies
are considered. Numerical results are presented which demonstrate
the sensitivity of the no-arbitrage fee to economic, contractual and
longevity assumptions. 

\vspace{1cm}


\vspace{1cm}
\vspace{1cm}

\newpage

\section{Introduction}

Variable Annuities are investment contracts with insurance coverage.
In recent years, they become a source of attraction for many investors,
because of their specific features: they are tax-deferral products
able to guarantee a minimum return for a long period, and to take
advantage of favorable market movements. The models in the literature,
present the analysis of these products in the Black and Scholes model,
disregarding possible changes in the interest rate and the volatility
of the underlying.

In 2008 following the subprime crisis, financial markets have suffered
the upheavals that have affected the entire world economy. Since then,
these markets were extremely volatile. After many failures, the gap
between the different interest rates applied to different transmitters
has become larger and larger, and a discussion on the identification
of the risk-free rate is open. The ECB and the Fed's interest rates
gradually declined, while the rate on sovereign debt increased gradually.

In this article, we consider a Guaranteed Minimum Withdrawal Benefit
(GMWB) annuity. We restrict our attention to a simplified form of
a GMWB which is initiated by making a lump sum payment to an insurance
company. This lump sum is then invested in risky assets, usually a
mutual fund. The benefit base, or guarantee account balance, is initially
set to the amount of the lump sum payment. The holder of the policy
(hereinafter, we will abbreviate it with \emph{PH}) is entitled to
withdraw a fixed sum, even if the actual investment in the risky asset
declines to zero. The withdrawal period may start immediately or later:
in this case the benefit base and the account value may be reset to
the maximum between their value and a fixed value. Finally, the PH
may withdraw more than the contractually specified amount, including
complete surrender of the contract, upon payment of a penalty. Complete
surrender here means that the PH withdraws the entire amount remaining
in the investment account, and the contract terminates. In most cases,
this penalty for full or partial surrender declines to zero after
five to seven years. During contract execution, a death benefit may
come with the PH's death: in this case, his (her) heirs receive the
remaining amount in the risky asset account.

The hedging costs for this guarantee are offset by deducting a proportional
fee from the risky asset account. From an insurance point of view,
these products are treated as financial ones: the products are hedged
as if they were pure financial products, and the mortality risk is
hedged using the law of large numbers. Therefore, it's very important
for insurance companies to be able to price quickly these products.
Moreover these products have long maturities that could last almost
25 years. The Black-Scholes model, with constant interest rate and
volatility seems to be unsuitable for those products: that\textquoteright{}s
why we present our pricing methods in two frameworks, modeling stochastic
volatility (Heston model \cite{He}) and stochastic interest rate
(Hull-White model \cite{HW}) . 

There have been several recent articles on pricing GMWBs. In particular,
we would remember Chen and Forsyth \cite{CF} and Chen, Vetzal and
Forsyth \cite{CVF}. In the first paper, the authors used an impulse
stochastic control formulation for pricing variable annuities GMWB,
assuming the PH to be allowed to withdraw funds continuously, or only
at anniversaries. In the second one, the authors analyzed the impact
of several product and model parameters using the same PDE approach.
The use of PDEs proved to be very fast and accurate, and we used it
as a reference for our work. 

Another research work about GMWB is Yang and Dai's one \cite{YD}:
they used a flexible tree for evaluating GMWB contracts with various
provisions. Yang-Dai's product is slightly different from Chen-Forsyth's
one: that's why we treat the two apart.

We have made reference also to Bacinello et al. \cite{bips}: variable
annuities (including GMWBs) are priced using a Monte Carlo approach.
The PH's behavior is assumed to be semi-static, i.e. the holder withdraws
at the contract rate or surrenders the contract.

In this paper, we price two types of GMWBs guarantees and we find
the no-arbitrage fee in the Heston model and the Black-Scholes with
stochastic interest rate model (\emph{BS HW model}). First, we treat
a static withdrawal strategy: the PH withdraws at the contract rate.
Then, taking the point of view of the worst case for the hedger, we
price the guarantees assuming that the PH follows a dynamic withdrawal
strategy. We also used these methods to calculate the Greeks for hedging
and risk management. For this purpose we present four numerical methods:
a hybrid tree-finite difference method and a Hybrid Monte Carlo method
(both introduced by Briani et al. \cite{BCZ0}) an ADI finite difference
scheme (Haentjens and Hout \cite{HH}),  and a Standard Monte Carlo
method with Longstaff-Schwartz least squares regression (Longstaff
and Schwartz \cite{LS}). These methods have already been developed
in the problem of ``Guaranteed Lifelong Withdrawal Benefit'' (GLWB)
pricing as we did in \cite{GMZ}, but in this case problem dimension
may change. In fact, GLWB pricing problem has dimension equal to 2,
while GMWB pricing problem may have dimension 2 or 3.

We use the term \emph{no-arbitrage fee} in the sense that this is
the fee which is required to maintain a replicating portfolio. A description
of the replicating portfolio for these types of guarantees is given
in Chen et al. \cite{CF} and Belanger et al. \cite{BF}. 

The main results of this paper are the following ones:
\begin{itemize}
\item We formulate the determination of the no-arbitrage fee (i.e. the cost
of maintaining a replicating hedging portfolio) in the Heston model
and in the BS HW model using different pricing methods;
\item We present the effects of stochastic volatility and stochastic interest
rate on pricing and Greeks calculation, and the sensitivity of the
GMWB fee to various modeling parameters;
\item We use different numerical methods to price the GMWB contracts;
\item We present numerical examples which show the convergence of these
methods.
\end{itemize}
The paper is organized as follows: in Section 2, we describe the main
features of the contracts such as event times, withdrawals and penalties.
In Section 3, we provide a brief review of the stochastic models used
afterward. In Section 4, we present the numerical methods, and how
to implement them to solve the GMWB contract pricing problem. In Section
5 we perform numerical tests in order to show their behavior and we
study the sensitivity of the no-arbitrage fee to economic and contractual
assumptions. Finally, in Section 6, we present the conclusions.

\section{\label{2}The GMWB Contracts}

In the following, we will refer to the contracts described in the
paper of Chen and Forsyth \cite{CF} and in the paper of Yang and
Dai \cite{YD}. We are calling GMWB-CF the contract described in \cite{CF}
and GMWB-YD the contract described in \cite{YD}. Now, we make a brief
summary of the main features of the two contracts.

\subsection{Mortality}

Similar to the work of Chen and Forsyth \cite{CF}, Milevsky and Salisbury
\cite{MS} and Dai et al. \cite{DKZ}, we will ignore mortality effects
in the following. We plan to study the effects of mortality in a future
work.

\subsection{Contract State Parameters }

At time $t=0$ the policy holder pays with lump sum the premium $P$
to the insurance company. The premium $P$ is invested in a fund whose
price is denoted by the variable $S_{t}$.

For both the two contracts, we suppose that there is a set of discrete
times $\left\{ t_{i},i=1,\dots N\right\} $, which we term \emph{event
times}; at these times withdrawals may occur. We suppose $\Delta t_{i}=t_{i+1}-t_{i}$
to be constant, and denoted by $\Delta t$. We also consider $t_{0}=t_{1}-\Delta t$,
and to be consistent with Yang Dai's notation we will write $T_{1}$
instead of $t_{0}$ ($T_{1}=t_{0}$). Then, we write $T_{1}<T_{1}+\Delta t=t_{1}<t_{2}<\dots<t_{N}=T_{2}$;
the time interval $\left[T_{1},T_{2}\right]$ is called \emph{payout
phase}. We remark that no withdrawals take place in $T_{1}$.

\subsubsection*{GMWB-CF}

The state parameters of the contract are:
\begin{itemize}
\item Account value: $A_{t}$, $A_{0}=P$.
\item Base benefit: $B_{t}$, $B_{0}=P$.
\end{itemize}
Both these two variables are initially set equal to the premium. We
define $T_{1}=0$ the time of the contract beginning, and $T_{2}=t_{N}$
the time of the last possible withdrawal. Usually, the first withdrawal
takes place in $t_{1}=1\ y$ or $t_{1}=0.5\, y$.

\subsubsection*{GMWB-YD}

The state parameters of the contract are:
\begin{itemize}
\item Account value: $A_{t}$, $A_{0}=P$.
\item Guaranteed minimum withdrawal: $G$.
\end{itemize}
The variable $A_{t}$ is initially set equal to the premium, while
$G$ is not defined until the beginning of the withdrawal period at
time $T_{1}$. For this type of contract we don't need to define the
Benefit Base variable because its value is deterministic until the
PH decides to lapse.

For this product, there exist two time parameters, $T_{1}$ and $T_{2}$
that express the begin and the end of the payout phase. Yang and Dai
used integers values for $T_{1}$ and $T_{2}$ and $\Delta t=1\, y$
in their numerical tests. No withdrawals happen during the deferred
time, i.e. for $t\in\left[0,T_{1}\right[$: in that period the account
value evolves as explained in the next subsection (see Formula (\ref{eq:fees_cont})).
At time $T_{1}$ also the account value is reset to
\[
A_{T_{1}^{\left(+\right)}}=\max\left[C\left(T_{1}\right),A_{T_{1}^{\left(-\right)}}\right],
\]
and the value of $G$ is fixed as 
\begin{equation}
G=\frac{A_{T_{1}^{\left(+\right)}}}{m\left(T_{2}-T_{1}\right)},\label{eq:YD-G}
\end{equation}

where $m$ denotes the number of withdrawals per year (usually $m=1$),
and $C\left(T_{1}\right)$ is a contract specified value that can
be interpreted as the lower bound of the total guaranteed withdrawal.
That value is specified as the return on the initial investment with
a roll-up interest rate guaranteed interest rate $i$, as follows:
\[
C\left(T_{1}\right)=P\left(1+i\right)^{T_{1}}.
\]

If $T_{1}=0$, the reset is trivial: $A_{T_{1}^{\left(+\right)}}=A_{0}=P$.

\subsection{Evolution of the Contracts in the Deferred Time and between Event
Times.}

We call \emph{deferred time} the time between $0$ and the beginning
of the payoff phase $T_{1}$: $0\leq t<T_{1}$. This time set is empty
unless for deferred GMWB-YD products; the other products have $T_{1}=0$
so there is no deferred time. We first consider the evolution of the
value of the guarantee excluding event times $t_{i}$. Let $t\in\left[0,T_{1}\right[\subseteq\left[0,T_{2}\right]$
or $t\in\left]t_{i},t_{i+1}\right[\subseteq\left[0,T_{2}\right]$.
As we said before, $S_{t}$ denotes the underlying fund driving the
account value. The dynamics of $S_{t}$ will be described in the next
Section. The account value $A_{t}$ follows the same dynamics of $S_{t}$
with the exception of the fact that some fees may be subtracted continuously:

\begin{equation}
dA_{t}=\frac{A_{t}}{S_{t}}dS_{t}-\alpha_{tot}A_{t}dt.\label{eq:fees_cont}
\end{equation}
We suppose that the total annual fees are charged to the PH and withdrawn
continuously from the investment account $A_{t}$. These fees include
the mutual fund management fees $\alpha_{m}$ and the fee charged
to fund the guarantee (also known as the rider) $\alpha_{g}$, so
that
\[
\alpha_{tot}=\alpha_{m}+\alpha_{g}.
\]
The only portion used by the insurance company to hedge the contract
is that coming from $\alpha_{g}$: the other part of the fees has
to be considered as an outgoing money flow as PH's withdrawals are.

\subsection{Event Times and Final Payoff}

Let $G$ be the withdrawal guaranteed amount: for a CF product type,
this parameter is a contract input, while for a YD type this value
is determined at time $T_{1}$ according to formula \ref{eq:YD-G}.

We denote $W_{i}$ the withdrawal of the PH at time $t_{i}$. As in
\cite{CF}, we observe that $W_{i}$ is a control variable.

\subsubsection*{GMWB-CF}

Usually the first event time takes place at time $t_{1}=\Delta t=1\, y$
or $t_{1}=\Delta t=0.5\, y$ ; moreover $t_{i}=i\cdot\Delta t$. 

We denote with $\left(A_{t_{i}^{\left(-\right)}},B_{t_{i}^{\left(-\right)}},t_{i}\right)$
the state variables just before an event time that occurs at time
$t_{i}$ and with $\left(A_{t_{i}^{\left(+\right)}},B_{t_{i}^{\left(+\right)}},t_{i}\right)$
the state variables just after it.

We distinguish two pricing frameworks: at each event time the PH can
withdraw according to the contract rate $G$ (\emph{Static approach})
or to a different rate (\emph{Dynamic approach}). If $W_{i}\leq G$,
then there is no penalty imposed; if $W_{i}>G$ there is a proportional
penalty charge $\kappa\left(W_{i}-G\right)$. Anyway, the value of
$W_{i}$ chosen by the PH cannot exceed the guaranteed withdrawal
amount $B_{t_{i}^{\left(-\right)}}$: it must be $W_{i}\in\left[0,B_{t_{i}^{\left(-\right)}}\right]$. 

As we said before, the PH may not receive all the money he (she) withdraws
from the account value. Let $f_{i}\left(W\right):\left[0,B_{t_{i}^{\left(-\right)}}\right]\rightarrow\mathbb{R}$
be a function of $W_{i}$ denoting the rate of cash flow received
by the PH due to the withdrawal at time $t_{i}$. Then, 
\[
f_{i}\left(W_{i}\right)=\begin{cases}
W_{i} & \mbox{if }W_{i}\leq G\\
W_{i}-\kappa\left(W_{i}-G\right) & \mbox{if }W_{i}>G.
\end{cases}
\]

The new state variables are
\begin{equation}
\left(A_{t_{i}^{\left(+\right)}},B_{t_{i}^{\left(+\right)}},t_{i}\right)=\left(\max\left(A_{t_{i}^{\left(-\right)}}-W_{i},0\right),B_{t_{i}^{\left(-\right)}}-W_{i},t_{i}\right).\label{eq:ET-CF}
\end{equation}

At time $t=T_{2}$ the last event time takes place: the PH withdraws
as in the previous event times; then he (she) receives the final payoff
which is worth
\[
FP=\max\left(A_{T_{2}},\left(1-\kappa\right)B_{T_{2}}\right).
\]

This final payoff is applied also in the static case. 

It is possible to prove that the optimal withdrawal at time $T_{2}$
is $W_{N}=\min\left(G,B_{T_{2}^{\left(-\right)}}\right)$; in this
case, the value of the contract before the withdrawal is 
\[
\mathcal{V}\left(A_{T_{2}^{\left(-\right)}},B_{T_{2}^{\left(-\right)}},T_{2}\right)=\max\left(A_{T_{2}^{\left(-\right)}},\left(1-\kappa\right)B_{T_{2}^{\left(-\right)}}+\kappa\min\left(G,B_{T_{2}^{\left(-\right)}}\right)\right).
\]
Therefore, this remark simplifies the research of the optimal withdrawal
in the Dynamic framework.

We notice that, if $A_{t_{i}^{\left(-\right)}}>B_{t_{i}^{\left(-\right)}}$
the contract can not be fully terminated in $t_{i}$: if the PH withdraws
at the maximal rate, then $W_{i}=B_{t_{i}^{\left(-\right)}}$ and
$\left(A_{t_{i}^{\left(+\right)}},B_{t_{i}^{\left(+\right)}},t_{i}\right)=\left(A_{t_{i}^{\left(-\right)}}-B_{t_{i}^{\left(-\right)}},0,t_{i}\right)$.
In this case the PH won't be able to make any withdrawal in following
event times because of $B=0$, but he will receive the final payoff
$FP=A_{T_{2}}$ at time $T_{2}$.

\subsubsection*{GMWB-YD}

This kind of products can be deferred or not. If we set $\Delta t=\nicefrac{\left(T_{2}-T_{1}\right)}{N}$,
then $t_{i}=T_{1}+\Delta t\cdot i$ for $i=1,\dots,N$. Usually $\Delta t=1\, y$.

We denote with $\left(A_{t_{i}^{\left(-\right)}},G^{\left(-\right)},t_{i}\right)$
the state variables just before an event time that occurs at time
$t_{i}$ and with $\left(A_{t_{i}^{\left(+\right)}},G^{\left(+\right)},t_{i}\right)$
the state variables just after it. 

According to \cite{YD}, we distinguish two pricing frameworks: at
each event time the PH can withdraw according to the contract rate
$G$ (\emph{Static approach}) or fully surrender (\emph{Dynamic approach}).
In the first case, he (she) receives $G$ at all event times after
$T_{1}$ ($T_{2}-T_{1}$ payments) and the state change is given by
\begin{equation}
\left(A_{t_{i}^{\left(+\right)}},G^{\left(+\right)},t_{i}\right)=\left(\max\left(0,A_{t_{i}^{\left(+\right)}}-G^{\left(-\right)}\right),G^{\left(-\right)},t_{i}\right).\label{eq:ET-YD}
\end{equation}

At time $t=T_{2}$, the PH receives $G$ plus the final payoff: 
\[
FP=A_{T_{2}^{\left(+\right)}}.
\]

In the second case, the PH receives $G$ until the surrender event,
and the equation (\ref{eq:ET-YD}) still holds. Let's suppose that
the PH decides to surrender at time the event time $t_{i^{*}}$; then
\[
\left(A_{t_{i^{*}}^{\left(+\right)}},G,t_{i^{*}}\right)=\left(0,0,t_{i^{*}}\right).
\]
 The final payoff is paid out at time $t_{i^{*}}$, and the contract
becomes valueless: 
\[
FP=G+\left(1-\kappa\right)\max\left(0,A_{t_{i^{*}}^{\left(-\right)}}-G\right).
\]

\subsection{\label{sr}Similarity Reduction}

An important property of GMWB-YD contract is the fact that this contract
behaves good under scaling transformations as also GLWB variable annuities
do. If $\mathcal{V}\left(A,G,t\right)$ denotes the value of a contract,
it is possible to prove that for any scalar $\eta>0$
\begin{equation}
\eta\mathcal{V}\left(A,G,t\right)=\mathcal{V}\left(\eta A,\eta G,t\right).\label{eq:similarity}
\end{equation}

Then, we just have to treat the case $G=\hat{G}$ for a fixed value
$\hat{G}$ (for example $\hat{G}=P/\left(T_{2}-T_{1}\right)$), and
then, choosing $\eta=\nicefrac{\hat{G}}{G}$, we can obtain 
\[
\mathcal{V}\left(A,G,t\right)=\frac{G}{\hat{G}}\mathcal{V}\left(\frac{\hat{G}}{G}A,\hat{G},t\right),
\]

which means that we can solve the pricing problem only for a single
representative value of $G$. This effectively reduces the problem
dimension. 

The previous property can be applied at time $T_{1}$ when $A$ and
$G$ are reset. Some simple calculations show that
\[
\mathcal{V}\left(A_{T_{1}^{\left(+\right)}},G_{T_{1}}^{+},T_{1}\right)=\frac{A_{T_{1}^{\left(+\right)}}}{P}\mathcal{V}\left(P,\hat{G},T_{1}\right).
\]

The similarity reduction (\ref{eq:similarity}) was also exploited
from Shah et Bertsimas in \cite{SB}. We would remark that Yang and
Dai didn't use this technique for their product: therefore, their
resolution of the problem of pricing is more complicated and computationally
expensive.

According to GMWB-CF contracts, the similarity reduction can't be
applied directly. In fact, we can prove that 
\begin{equation}
\eta\mathcal{V}\left(A,B,G,t\right)=\mathcal{V}\left(\eta A,\eta B,\eta G,t\right)
\end{equation}

but in this case we have to scale also the guaranteed withdrawal amount
$G$ and therefore it is not useful to reduce problem dimension.

\section{\label{3}The Stochastic Models of the Fund $S$}

To understand the different impacts of stochastic volatility and stochastic
interest rate over such a long maturity contract, we price the GMWB
VA according to two models: the Heston model, which provides stochastic
volatility, and  the Black-Scholes Hull-White model, which provide
stochastic interest rate. As we said before, the process $S$ represents
the underlying fund driving the account value $A_{t}$ of the product.

\subsection{The Heston Model}

The Heston model \cite{He} is one of the most known and used models
in finance to describe the evolution of the volatility of an underlying
asset and the underlying asset itself. In order to fix the notation,
we report its dynamics:
\begin{equation}
\begin{cases}
dS_{t}=rS_{t}dt+\sqrt{v_{t}}S_{t}dZ_{t}^{S} & S_{0}=\bar{S}_{0},\\
dv_{t}=k\left(\theta-v_{t}\right)dt+\omega\sqrt{v_{t}}dZ_{t}^{v} & v_{0}=\bar{v}_{0},
\end{cases}\label{eq:He}
\end{equation}
where $Z^{S}$ and $Z^{v}$ are Brownian motions, and $d\left\langle Z_{t}^{S},Z_{t}^{v}\right\rangle =\rho dt$.

\subsection{The Black-Scholes Hull-White Model}

The Hull-White model \cite{HW} is one of historically most important
interest rate models, which is nowadays often used for risk-management
purposes. The important advantage of the HW model is the existence
of closed formulas to calculate the prices of bonds, caplets and swaptions.
In order to fix the notation, we report the dynamics of the BS HW
model:
\[
\begin{cases}
dS_{t}=r_{t}S_{t}dt+\sigma S_{t}dZ_{t}^{S} & S_{0}=\bar{S}_{0},\\
dr_{t}=k\left(\theta_{t}-r_{t}\right)dt+\omega dZ_{t}^{r} & r_{0}=\bar{r}_{0},
\end{cases}
\]
where $Z^{S}$ and $Z^{r}$ are Brownian motions, and $d\left\langle Z_{t}^{S},Z_{t}^{r}\right\rangle =\rho dt$. 

The process $r$ is a generalized Ornstein-Uhlenbeck (hereafter OU)
process: here $\theta_{t}$ is not constant but it is a deterministic
function which is completely determined by the market values of the
zero-coupon bonds (\emph{ZCBs}) by calibration (see Brigo and Mercurio
\cite{BM}): in this case the theoretical prices of the ZCBs match
exactly the market prices.

Let $P^{M}\left(0,T\right)$ denote the market price of the ZCB at
time $0$ for the maturity $T$. The market instantaneous forward
interest rate is then defined by
\[
f^{M}\left(0,T\right)=-\frac{\partial\ln P^{M}\left(0,T\right)}{\partial T}.
\]
It is well known that the short rate process $r$ can be written as
\[
r_{t}=\omega X_{t}+\beta\left(t\right),
\]
where $X$  is a stochastic process given by 
\[
dX_{t}=-kX_{t}dt+dZ_{t}^{r},\ X_{0}=0,
\]
and $\beta\left(t\right)$ is a function
\[
\beta\left(t\right)=f^{M}\left(0,t\right)+\frac{\omega^{2}}{2k^{2}}\left(1-\exp\left(-kt\right)\right)^{2}.
\]
Then, the BS HW model is described by
\begin{equation}
\begin{cases}
dS_{t}=r_{t}S_{t}dt+\sigma S_{t}dZ_{t}^{S} & S_{0}=\bar{S}_{0},\\
dX_{t}=-kX_{t}dt+dZ_{t}^{r} & X_{0}=0,\\
r_{t}=\omega X_{t}+\beta\left(t\right).
\end{cases}\label{eq:HW}
\end{equation}

A particular case is called \emph{flat curve}. In this case, we assume
$P^{M}\left(t,T\right)=e^{-\bar{r}_{0}\left(T-t\right)}$ and $f^{M}\left(0,T\right)=\bar{r}_{0}$.
Then 
\[
\beta\left(t\right)=\bar{r}_{0}+\frac{\omega^{2}}{2k^{2}}\left(1-\exp\left(-kt\right)\right)^{2},
\]
 and

\[
\theta_{t}=\bar{r}_{0}+\frac{\omega^{2}}{2k^{2}}\left(1-\exp\left(-2kt\right)\right).
\]

\section{\label{4}Numerical Methods of Pricing}

In this Section we describe the four pricing methods: a Hybrid Monte
Carlo method, a Standard Monte Carlo method, a Hybrid PDE method,
and an ADI PDE method. 

We remember that our aim is to find the fair value for $\alpha_{g}$:
it's the charge that makes the initial value of the policy equal to
the initial premium. To achieve this target, we price the policy (with
one of the following procedures) and then we use the secant method
to approach the correct value for $\alpha_{g}$. Therefore, the main
goal is to be able to find the initial value for a given value of
$\alpha_{g}$: $\mathcal{V}\left(A_{0},B_{0},0\right)\left(\alpha_{g}\right)$.

We remark that we want to calculate the value of the policy from the
point of view of the insurance company: the management fees are treated
as a outgoing cash flows, and if we assume that the policy holder
follows a withdrawal strategy, we consider the worst one for the insurance
company.

\subsection{The Hybrid Monte Carlo Method}

The value of a GMWB policy can be calculated through a Monte Carlo
set of simulations. This procedure is based in two steps: generation
of a scenario (a sampling of the underlying values along the life
of the product), and projection of the product into the scenario.
According to the way we obtain the scenarios, we distinguish two Monte
Carlo models: Hybrid MC (HMC) and Standard MC (SMC).

The Hybrid MC method was introduced in Briani et al. \cite{BCZ}.
It is a simple and efficient way to produce MC scenarios for different
models. This method is called ``hybrid'' because it combines trees
and MC methods. First, a simple  tree needs to be built: this can
be done according to Appolloni et al. \cite{AC} or, as we did for
GLWB in \cite{GMZ}. Then, using a vector of Bernoulli random variables,
we move from the root through the tree, describing the scenario for
the volatility or the interest rate. The values of the underlying
at each time step can be easily obtained using an Euler scheme.

\subsubsection{\label{tree}Trees}

\begin{figure}
\begin{centering}
\includegraphics[scale=0.7]{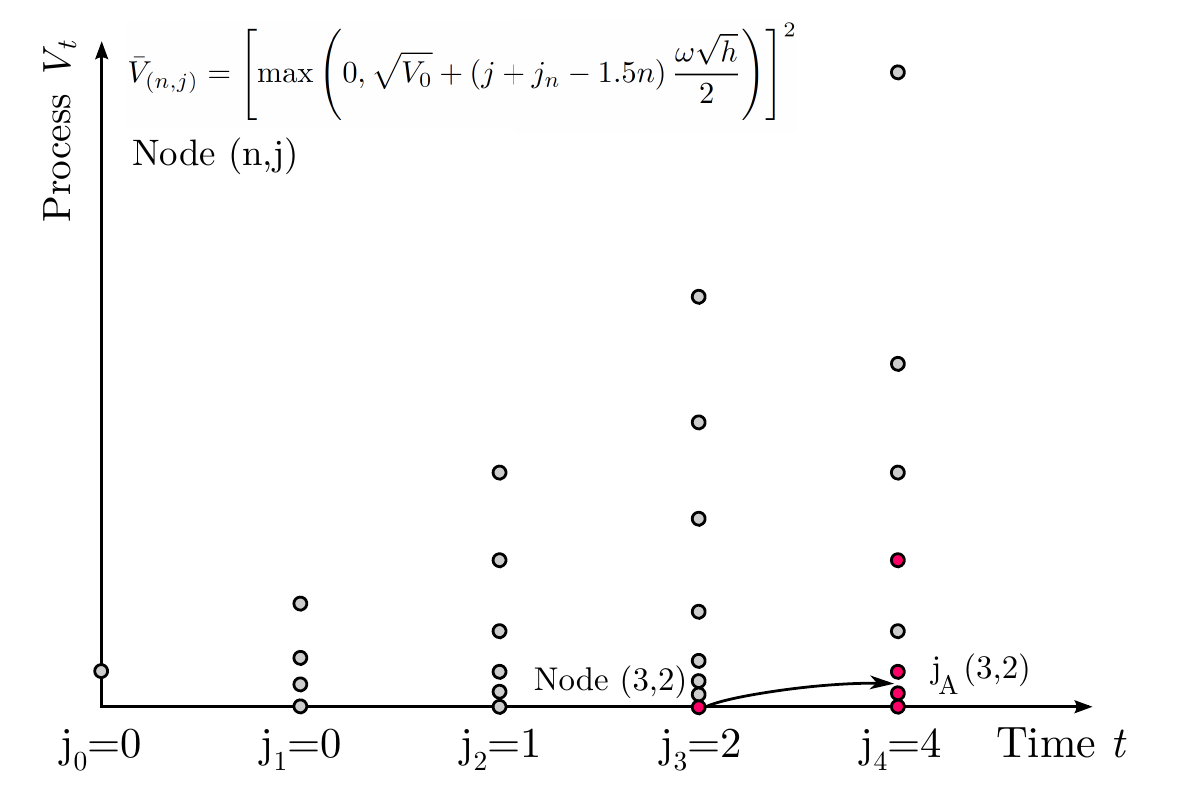}\includegraphics[scale=0.7]{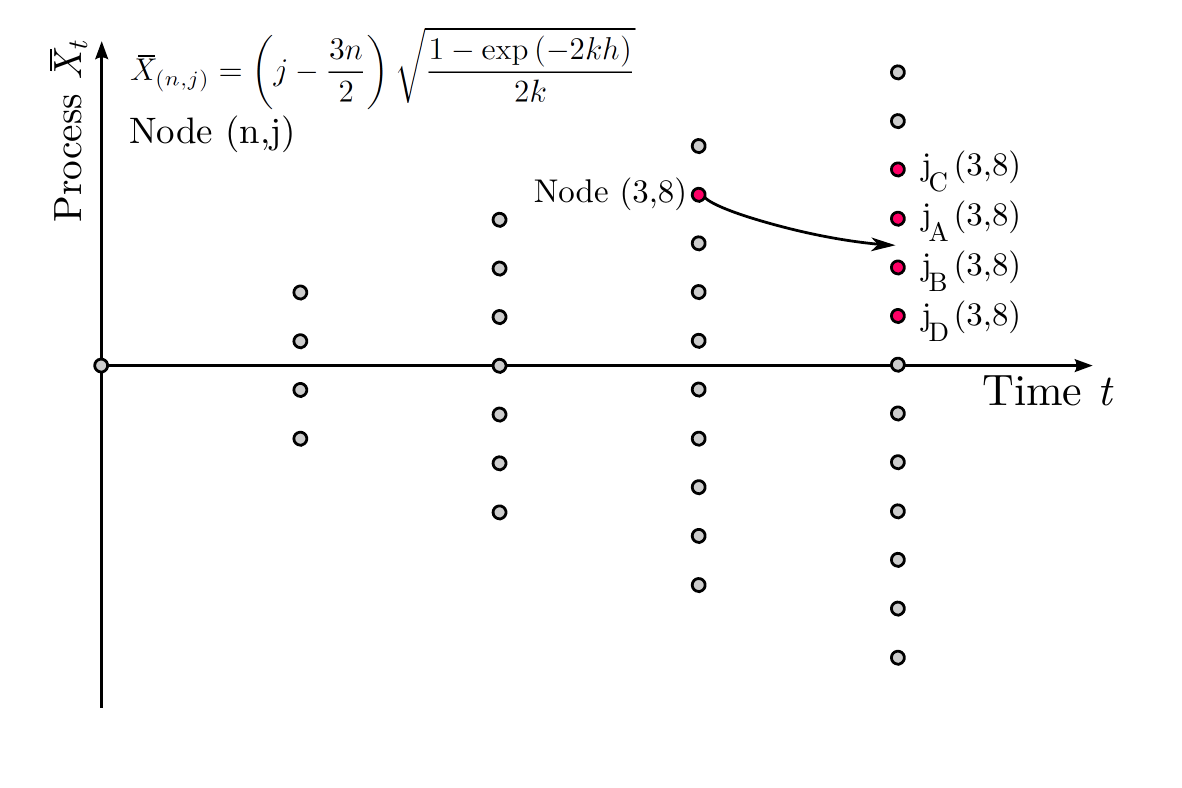}
\par\end{centering}

\caption{\label{fig:Trees} The trees for Heston and Hull-White models.}
\end{figure}

In this Subsection, we present the main points useful to build a tree
for volatility or interest rate: they're simple quadrinomial trees,
built to match the first 3 moments of the stochastic processes. Other
trees can be built according to Appolloni et al. \cite{AC} or Nelson
and Ramaswamy  \cite{NR}, but we preferred our quadrinomial trees
because they give more guarantee of convergence to the exact price
also for long maturities. An example of those trees is available in
Figure \ref{fig:Trees}.

We suppose to fix a time horizon $\left[0,T\right]$, a number $N>0$
and we define $h=\nicefrac{T}{N}$. We denote $\bar{G}_{\left(n,j\right)}$
the value of a node of tree level $n$ and position $j$.

\paragraph*{The Heston Model}

The Heston process (\ref{eq:He}) for volatility has no constant variance
and isn't Gaussian. We consider the process obtained by the square
root:
\[
d\sqrt{v_{t}}=\frac{4k\left(\theta-\sqrt{v_{t}}^{2}\right)-\omega^{2}}{8\sqrt{v_{t}}}dt+\frac{\omega}{2}dZ_{t}.
\]

We approximate it with a Gaussian process with variance $\frac{\omega^{2}}{4}dt$.
This approximation is helpful to define the grid of states-space for
the Markov chain: inspired by \cite{NR}, we define 
\[
j_{n}=\max\left(0,\mbox{floor}\left(1.5n-\frac{2\sqrt{v_{0}}}{\omega\sqrt{h}}\right)\right),
\]

and we set 
\[
\bar{v}_{\left(n,j\right)}=\left(\max\left(0,\sqrt{v_{0}}+\left(j+j_{n}-1.5n\right)\frac{\omega\sqrt{h}}{2}\right)\right)^{2}.
\]

for $j=0,\dots,3n-j_{n}$. The shift due to $j_{n}$ helps to reject
the many nodes with value equal to zero: if $j_{n}>0$, then $\bar{v}_{\left(n,0\right)}=0$
and $\bar{v}_{\left(n,1\right)}>0$ .

We would remark that the process $\bar{v}_{t}$ approximates $v_{t}$
and not $\sqrt{v_{t}}$: the moments matching is done according to
the moments of the process $v_{t}$.

We fix now the values of $n$ and $j$. The discrete process $\bar{v}$
can jump from a node to another, as in a Markovian chain. We show
now how to find the possible upcoming nodes.

The first three moments for the Heston process can be found in Alfonsi
\cite{AA}:
\[
\psi\left(h\right)=\nicefrac{\left(1-e^{-kh}\right)}{k},\ \ \ M_{1}=\mathbb{E}\left[v_{t+h}|v_{t}=v\right]=ve^{-kh}+\theta k\psi\left(h\right),
\]

\[
M_{2}=\mathbb{E}\left[\left(v_{t+h}\right)^{2}|v_{t}=v\right]=M_{1}^{2}+\omega^{2}\psi\left(h\right)\left[\theta k\psi\left(h\right)/2+ve^{-kh}\right],
\]

\[
M_{3}=\mathbb{E}\left[\left(v_{t+h}\right)^{3}|v_{t}=v\right]=M_{1}M_{2}+\omega^{2}\psi\left(h\right)\left[2v^{2}e^{-2kh}+\psi\left(h\right)\left(k\theta+\frac{\omega^{2}}{2}\right)\left(3ve^{-kh}+\theta k\psi\left(h\right)\right)\right].
\]

\begin{figure}
\begin{centering}
\includegraphics{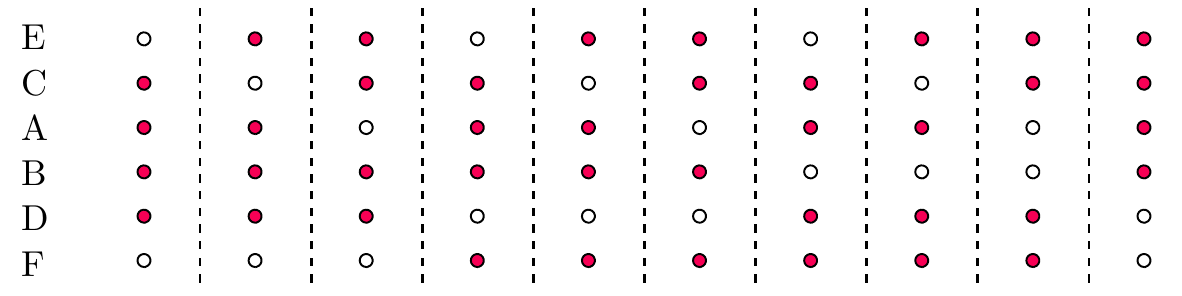}
\par\end{centering}

\caption{\label{fig:casi} The possible combinations used to get positive probabilities
in the Heston model tree. The red points correspond to the used nodes.}
\end{figure}

Then, we can proceed as in the general case. Anyway, the grid we're
using is based on an approximation: so the probabilities obtained
solving the linear system may not be positive. If we get a negative
transition probability for a given node, we try another combination
of out-coming nodes, replacing one (or two) of the nodes $A$, $B$,
$C$, $D$ with one (or two) close to them. The nodes $A$ or $C$
may be replaced by a node $E$ defined as the first node bigger than
$C$, and the nodes $B$ or $D$ may be replaced with a node $F$,
defined as the smallest before node $D$. This gives rise to $9$
combinations to be tested. If the starting node is small and the node
$D$ verifies $j_{D}=j_{n}$ we could not do this last change because
there would be no $F$ node. In this case we allow the node $D$ to
be replaced by the node $E$: see Figure \ref{fig:casi}.

If these attempts don't give a positive result (negative probabilities),
we give up trying to match the first three moments, and we are content
to match an approximation of the first two as in \cite{NR}, thus
ensuring the weak convergence. In this case, we only use the nodes
$A,B,C,D$: we define 
\[
p_{AB}=\frac{\mu-G_{n+1,j_{B}}}{G_{n+1,j_{A}}-G_{n+1,j_{B}}},\ \ p_{BA}=1-p_{AB},
\]
\[
p_{CD}=\frac{\mu-G_{n+1,j_{D}}}{G_{n+1,j_{C}}-G_{n+1,j_{D}}},\ \ p_{DC}=1-p_{CD},
\]

and 
\[
p_{A}=\frac{5}{8}p_{AB},\ p_{B}=\frac{5}{8}p_{BA},\ p_{C}=\frac{3}{8}p_{CD},\ p_{D}=\frac{3}{8}p_{DC}.
\]

It is possible to show that the first moment of this variable is equal
to $M_{1}$, and as $h\rightarrow0$ the second moment approaches
to $M_{2}$, ensuring the convergence, as proved in \cite{NR}.

In all our numerical tests, this last option (matching only two moments)
has never been necessary: changing the nodes, all moments were matched
with positive probabilities.

\paragraph*{The Hull-White Model}

The process $X$ in (\ref{eq:HW}) is Gaussian, and this property
simplifies the tree construction. As shown in Ostrovski \cite{VO}
the variables $X_{t}$ and $\int_{s}^{t}X_{y}dy$ are bivariate normal
distributed conditionally on $X_{s}$ with well known mean and variance.
We define 
\[
X_{\left(n,j\right)}=\left(j-\frac{3n}{2}\right)\sqrt{\frac{1-\exp\left(-2kh\right)}{2k}},\ n=0,\dots,N\mbox{ and }j=0,\dots,3n.
\]

Let's fix a node $X_{\left(n,j\right)}$. We define
\[
H=\exp\left(-kh\right),\ K=\sqrt{\frac{1-\exp\left(-2kh\right)}{2k}},\ M_{1}=X{}_{\left(n,j\right)}H,
\]
 
\[
j_{A}=\mbox{ceil}\left[\frac{M_{1}}{K}+\frac{3\left(n+1\right)}{2}\right],\ X_{A}=X_{\left(n+1,j_{A}\right)}.
\]

The transition probabilities are given by
\[
\begin{array}{cc}
p_{A}=\frac{\left(X_{A}-M_{1}\right)}{2K^{3}}\left(K^{2}+\left(K+M_{1}-X_{A}\right)^{2}\right), & p_{B}=\frac{\left(K+M_{1}-X_{A}\right)}{2K^{3}}\left(K^{2}+\left(M_{1}-X_{A}\right)^{2}\right),\\
p_{C}=\frac{\left(K+M_{1}-X_{A}\right)}{6K^{3}}\left(2K^{2}+\left(K+M_{1}-X_{A}\right)^{2}\right), & p_{D}=\frac{\left(X_{A}-M_{1}\right)}{6K^{3}}\left(2K^{2}+\left(M_{1}-X_{A}\right)^{2}\right).
\end{array}
\]

\subsubsection{Scenario Generation}

The generations of the volatility process and of the interest rate
process behaves in a similar way: we start from the node $\left(0,0\right)$
of the tree and according to a discrete random variable and to the
node probabilities, we move to the next node and so on. Let $R$ be
a discrete random variable that can assume value $A,B,C,D$ with probabilities
$p_{A},p_{B},p_{C},p_{D}$: sampling such a variable at each node,
we get the values of the process at each time step.

We distinguish two cases for the two models.

\paragraph*{The Heston Model}

We approximate the couple $\left(S_{t},v_{t}\right)$ in $\left[0,T\right]$
by a discrete process $\left(\bar{S}_{k\Delta t},\bar{v}_{k\Delta t}\right)_{k=0,\dots,T/\Delta t}$,
with $\left(\bar{S}_{0},\bar{v}_{0}\right)=\left(S_{0},v_{0}\right)$.
For each scenario, we generate the volatility. 

Let $N\sim\mathcal{N}\left(0,1\right)$ and $B\sim\mathcal{B}\left(0.5\right)$.
We deduce the value of $\bar{S}_{t+\Delta t}$ by 
\[
\bar{S}_{t+\Delta t}=\begin{cases}
\bar{S}_{t}\exp\!\left[\!\left(r-\frac{\rho}{\sigma}k\theta\right)\!\Delta t\!+\!\left(\frac{\rho}{\sigma}k-\frac{1}{2}\right)\!\!\left(\!\frac{\bar{v}_{t+\Delta t}+\bar{v}_{t}}{2}\!\right)\!\Delta t\!+\!\frac{\rho}{\sigma}\!\left(\bar{v}_{t+\Delta t}-\bar{v}_{t}\right)\!+\!\sqrt{\!\left(1-\rho^{2}\right)\Delta t\bar{v}_{t}}N\!\right] & \mbox{if }B=0,\\
\bar{S}_{t}\exp\!\left[\!\left(r-\frac{\rho}{\sigma}k\theta\right)\!\Delta t\!+\!\left(\frac{\rho}{\sigma}k-\frac{1}{2}\right)\!\!\left(\!\frac{\bar{v}_{t+\Delta t}+\bar{v}_{t}}{2}\!\right)\!\Delta t\!+\!\frac{\rho}{\sigma}\!\left(\bar{v}_{t+\Delta t}-\bar{v}_{t}\right)\!+\!\sqrt{\!\left(1-\rho^{2}\right)\Delta t\bar{v}_{t+\Delta t}}N\!\right] & \mbox{if }B=1.
\end{cases}
\]

According to (\ref{eq:He}), we use the normal variable $N$ to generate
the Gaussian increment of $S$, and the Bernoulli variable $B$ to
split the operator associated to the Heston process. This scheme (without
splitting) appears in Briani et al. \cite{BCZ} and the splitting
method appears in Alfonsi \cite{AA}.

\paragraph*{The Black-Scholes Hull-White Model}

We approximate the couple $\left(S_{t},X_{t}\right)$ in $\left[0,T\right]$
by a discrete process $\left(\bar{S}_{k\Delta t},\bar{X}_{k\Delta t}\right)_{k=0,\dots,T/\Delta t}$,
with $\left(\bar{S}_{0},\bar{X}_{0}\right)=\left(S_{0},0\right)$,
and we deduce the interest rate by $\bar{r}_{t}=\omega\bar{X}_{t}+\beta\left(t\right)$.
Let $N\sim\mathcal{N}\left(0,1\right)$. We deduce the value of $\bar{S}_{t+\Delta t}$
by
\[
\bar{S}_{t+\Delta t}=\bar{S}_{t}\exp\left[\left(\frac{\bar{r}_{t\Delta t}+\bar{r}_{t}}{2}-\frac{\sigma^{2}}{2}\right)\Delta t+\sigma\left(\left(\bar{X}_{t+\Delta t}+\bar{X}_{t}\left(k\Delta t-1\right)\right)\rho+\sqrt{\Delta t}\bar{\rho}N\right)\right].
\]

\subsubsection{\label{Proj}Projection}

Once we have generated the scenarios set $\mathcal{S}=\left\{ s_{k},k=1,\dots,n_{s}\right\} $,
we project the policy into all of it's scenarios: this means we calculate
the initial value $V_{s}$ of the contract as the sum of discounted
cash flows determined according to each scenario $s\in\mathcal{S}$.
Then, the initial value of the contract $\mathcal{V}$ can be approximates
as the average of the initial values among all scenarios: 
\[
\mathcal{V}\approx\sum_{k=1}^{n_{s}}\frac{V_{s_{k}}}{n_{s}}.
\]
This calculation depends on whether we take an optimized strategy
or not.

\paragraph*{Constant Withdrawal}

In this case the strategy of the PH is fixed. We set 
\[
G=\frac{P}{m\left(T_{2}-T_{1}\right)}.
\]
 For a GMWB-CF product, the value of the base benefit $B_{t_{i}}$
is certain: $B_{t_{i}^{\left(-\right)}}=P-G\left(i-1\right)$ and
$B_{t_{i}^{\left(+\right)}}=B_{t_{i}^{\left(-\right)}}-G$. We can
just write $V_{s}=V_{s}\left(A,t\right)$ to denote the value of the
GMWB having account value equal to $A$ at time $t$. This fact sets
the problem dimension to 2. In this case, GMWB-CF and GMWB-YD collapse
in the same product.

For each scenarios $s$, first we calculate the values $A_{t_{i}^{\left(+\right)}}$
for all $t_{i}$:
\[
\begin{cases}
A_{T_{1}}=P\\
A_{t_{i}^{\left(-\right)}}=A_{t_{i-1}^{\left(+\right)}}\frac{S_{t_{i}}}{S_{t_{i-1}}}e^{-\alpha_{tot}\Delta t}\\
A_{t_{i}^{\left(+\right)}}=\max\left(0,A_{t_{i}^{\left(-\right)}}-G\right).
\end{cases}
\]

Then we set 
\[
V_{s}\left(A_{T_{2}^{\left(+\right)}},T_{2}\right)=A_{T_{2}^{\left(+\right)}};
\]

for all $T_{1}<t_{i}<T_{2}$ we have
\[
V_{s}\left(A_{t_{i}^{\left(+\right)}},t_{i}\right)=e^{-\int_{t_{i}}^{t_{i+1}}r_{s}ds}\left[V_{s}\left(A_{t_{i+1}^{\left(+\right)}},t_{i+1}\right)+G\right],
\]

and finally 
\[
V_{s}\left(A_{T_{1}},T_{1}\right)=e^{-\int_{T_{1}}^{t_{1}}r_{s}ds}\left[V_{s}\left(A_{t_{1}^{\left(+\right)}},t_{1}\right)+G\right].
\]

If $T_{1}=0$, then this is the initial value of the policy according
to the scenario $s$. Otherwise, if we're price a deferred product,
(i.e. $T_{1}>0$) we use similarity reduction to obtain
\[
V_{s}\left(A_{0},0\right)=e^{-\int_{0}^{T_{1}}r_{s}ds}V_{s}\left(P,T_{1}\right)\cdot\frac{\max\left(P,S_{T_{1}}e^{-\alpha_{tot}T_{1}}\right)}{P}.
\]

\paragraph*{Optimal Withdrawal}

The Optimal Withdrawal is a case of Dynamic Withdrawal and it applies
only to GMWB-CF product. In their articles, Chen and Forsyth suppose
the PH to be entitled to do optimal withdrawals, i.e. chose at each
event time how much withdraw. In this case we suppose that at each
event time $t_{i}$ the PH can withdraw a fraction of the regular
amount. To price in this case, we suppose that the PH chooses the
value of $W_{i}$ that causes the worst hedging case for the insurance
company. In this case, we denote $\mathcal{V}\left(A,B,t\right)$
the expected value at time $t$ of a generic policy whose state parameters
are $A,B$ :
\[
\mathcal{V}\left(A,B,t\right)=\mathbb{E}\left[V_{s}\left(A,B,t\right)\right].
\]
So, we suppose that the PH chooses $W_{i}$ such that 
\[
W_{i}=\underset{w_{i}\in\left[0,B_{t_{i}^{\left(-\right)}}\right]}{\mbox{argmax}}\ \mathcal{V}\left(\max\left(A_{t_{i}^{\left(-\right)}}-w_{i},0\right),B_{t_{i}^{\left(-\right)}}-w_{i},t_{i}\right)+f_{i}\left(w_{i}\right).
\]
This expected value can be calculated with a Longstaff-Schwartz approach:
\begin{enumerate}
\item Simulate $N$ random scenarios and price the policies into these scenarios.
\item For $i=N$ to $i=0$ (from $t_{N}=T_{2}$ to $t_{0}=T_{1}=0$):

\begin{enumerate}
\item For each scenario $s$ calculate $V_{s}\left(A_{t_{i}^{\left(+\right)}},B_{t_{i}^{\left(+\right)}},t_{i}\right)$
as sum of future discounted cash flows.
\item Approximate the function $\mathcal{V}\left(A,B,t_{i}\right)$ using
the least squares projection into a space of functions (usually polynomials)
(if $t_{i}>0$).
\item For each scenario $s$ find the optimal withdrawal $W_{i}$ (if $t_{i}>0$).
\item Recalculate the upcoming state variables from $\tau=t_{i}$ to $\tau=T_{2}$
assuming that the PH chooses the best value for $W_{\tau}$ (if $t_{i}>0$).
\end{enumerate}
\item Approximate $\mathcal{V}\left(P,P,0\right)$ with the average of the
initial values $\left\{ V_{s}\left(P,P,0\right),s\in\mathcal{S}\right\} $
.
\end{enumerate}
The search for the optimal withdrawal for this type of product is
a stiff purpose. The approximation of the function $\mathcal{V}\left(A,B,t\right)$
with polynomials is hard: this is due to the fact that this function
is very curved when the account value $A_{t}$ is close to $B_{t}$,
and is very straight otherwise. We developed the projection algorithm
in two different ways, to improve the computational time or the convergence
to the right value. We call the fast algorithm ``Full Regression''
and the accurate one ``Regression by Lines''.

\subparagraph{Full Regression}

\begin{figure}
\begin{centering}
\includegraphics[scale=0.58]{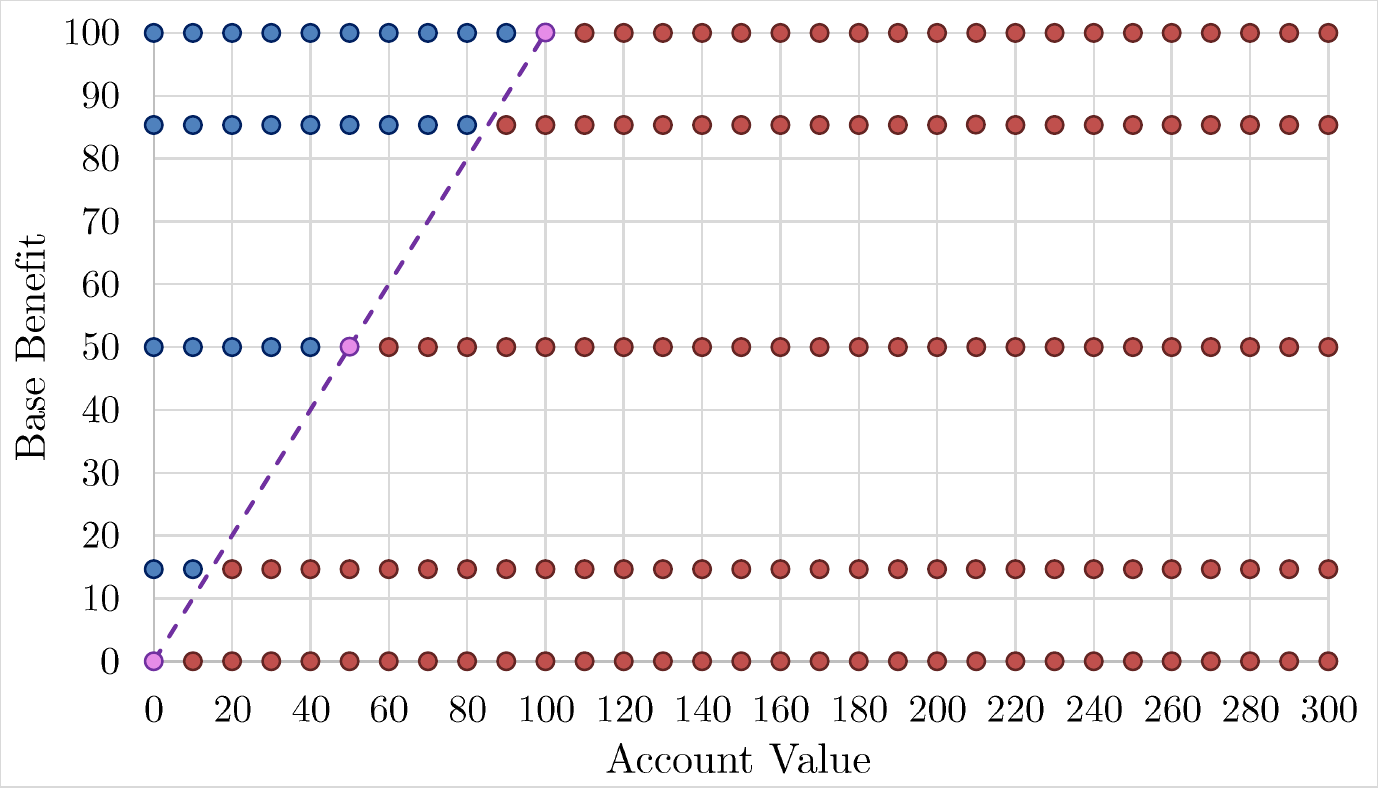}\hspace{0.1cm}\includegraphics[scale=0.58]{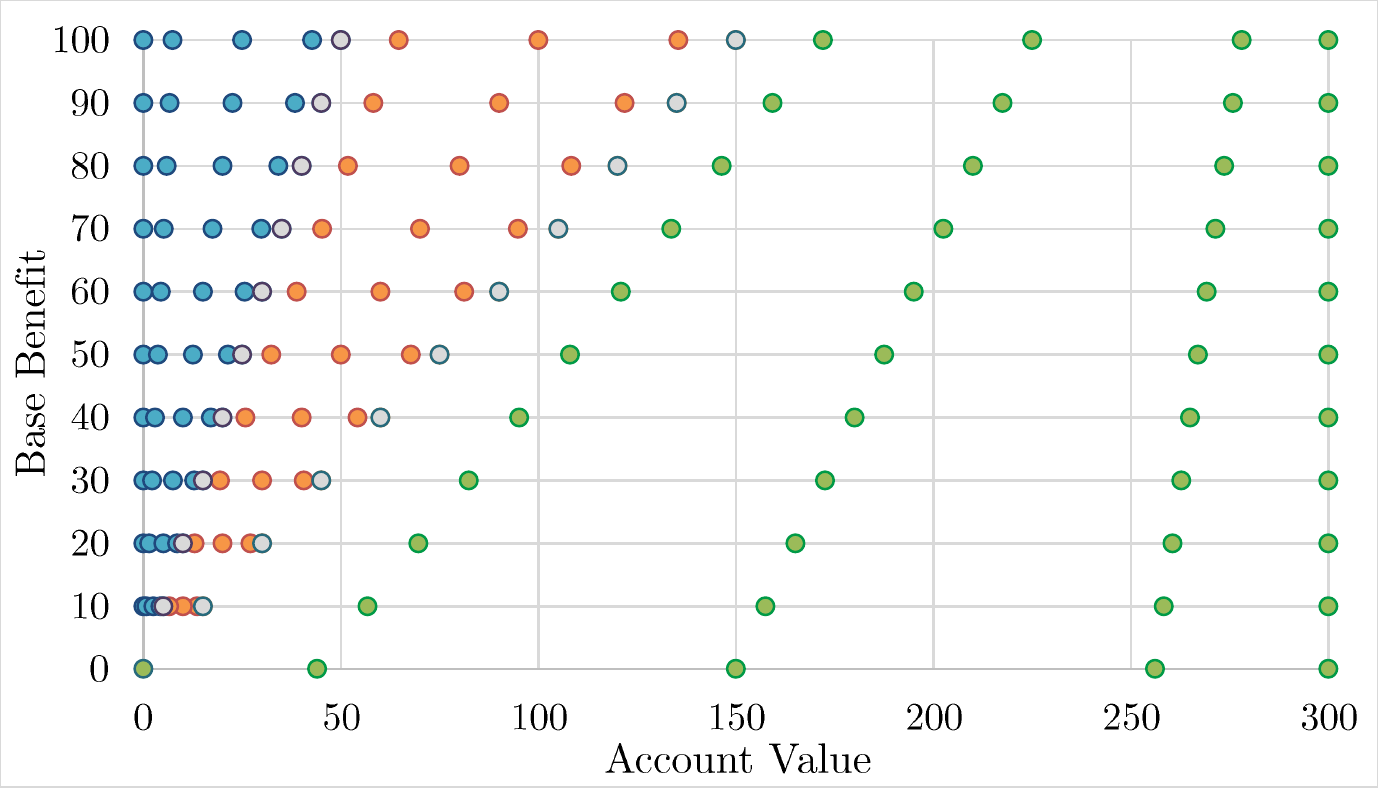}
\par\end{centering}

\caption{\label{fig:The-grids}The grids used in the Full Regression method
and in the Regression by Lines for a GMWB with $T_{2}=10$ and annual
withdrawals. In the first picture, violet points are used in both
the two regions. In the second picture, for each $B$ level, the gray
points border the different sectors and they are shared. The degree
of the polynomials for this last case is $4$.}
\end{figure}

In this case, the regression at each event time $t_{i}$ is done using
two polynomials with 3 variates: $Q_{t_{i}}^{up}\left(A,B,u\right)$
and $Q_{t_{i}}^{dw}\left(A,B,u\right)$ where $u$ is $r$ in the
BS HW model and $v$ in the Heston model. Here the most important
remarks
\begin{itemize}
\item Create a grid of constant points $\mathcal{G}=\mathcal{A}\times\mathcal{B}=\left\{ \left(a_{k},b_{h}\right),0\leq k\leq K,0\leq h\leq H\right\} $
to be used as initial values to diffuse the couple $\left(A,B\right)$
using random scenarios. This grid lets us be sure that at each event
time, the set of initial values is well distributed and useful for
polynomial regression. In our tests we used $\mathcal{B}$ as a set
of Chebychev nodes from $0$ to $P$, and $\mathcal{A}$ as a set
of uniform nodes from $0$ to $3P$. See Figure \ref{fig:The-grids}.
\item Separate the space in two regions $\mathcal{U}=\left\{ \left(a,b\right)|a\geq b\right\} $
and $\mathcal{D}=\left\{ \left(a,b\right)|a<b\right\} $ and perform
regression using $Q_{t_{i}}^{up}\left(A_{t_{i}},B_{tt_{i}},u_{t_{i}}\right)$
for the first set, and $Q_{t_{i}}^{dw}\left(A_{t_{i}},B_{t_{i}},u_{t_{i}}\right)$
for the second.
\item Use shift and scaling technique to improve regression.
\item As remarked before, the optimal withdrawal at last event time $t_{i}=T_{2}$
is always equal to $\min\left(G,B_{T_{2}^{\left(-\right)}}\right)$.
\item To find the best value for the withdrawal amount $W_{i}$, numerical
tests proved that, if $G$ divides exactly $P$, then it's enough
to search among the multiples of $G$.
\end{itemize}
Here a pseudo code:

\begin{Verbatim}[frame=single,fontsize=\footnotesize,numbers=left]
Full_regression(){
	int ETs= T2*WD_rate; 
	Scenario_generation_step();
	Forward_initial_step(); 	
	for(int ti= ETs-1;ti>0;ti--){ 		 		  
		Backward_step_GMWB(ti);  		 		
		Least_Squares_step_GMWB(ti);
		Forward_Dynamic_step_GMWB(ti); 
	}   
	Backward_step_GMWB(0); 
}
\end{Verbatim}

The functions that we used are the following ones:
\begin{itemize}
\item Scenario\_generation\_step(). Generate the scenarios: $S$ and $v$
or $r$ .
\item Forward\_initial\_step(). For all the scenarios $s$, chose a node
$\left(a,b\right)$ of the grid $\mathcal{G}$ (covering all the grid
as $s$ changes), and set $\left(A_{t_{i}}^{s},B_{t_{i}}^{s}\right)=\left(a,b\right)$
for all $t_{i}$. 
\item Backward\_step\_GMWB(ti). For all the scenarios $s$, calculate the
value of the policy $V_{t_{i}}^{s}$ at the event time $t_{i}$ as
the sum of discounted future cash flows.
\item Least\_Squares\_step\_GMWB(ti). Perform polynomial regression. Calculate
$Q_{t_{i}}^{up}\left(A,B,u\right)$ using the value $\left(A_{t_{i}},B_{t_{i}},u_{t_{i}},V_{t_{i}}\right)$
such that $A_{t_{i}}\geq B_{t_{i}}$. Calculate $Q_{t_{i}}^{dw}\left(A,B,u\right)$
using the value $\left(A_{t_{i}},B_{t_{i}},u_{t_{i}},V_{t_{i}}\right)$
such that $A_{t_{i}}\leq B_{t_{i}}$.
\item Forward\_Dynamic\_step\_GMWB(ti). Keeping fixed the value of $A_{t_{i}}$
and $B_{t_{i}}$ as stated by the Forward\_initial\_step function,
calculate the state parameters of the policy at all the event times
$t_{j}$ after $t_{i}$, using at each time the best withdrawal. As
we are proceeding backward and $t_{j}\geq t_{i}$, we can find the
best withdrawal using the polynomial $Q_{t_{j}}^{up}\left(A,B,u\right)$
and $Q_{t_{j}}^{dw}\left(A,B,u\right)$ calculated at the previous
steps. 
\end{itemize}

\subparagraph{Regression by Lines}

In this case, the regression at each event time $t_{i}$ is done using
3 polynomials with 2 variates for each value of base benefit $B$
and event time $t_{i}$: $Q_{t_{i},B}^{up}\left(A,u\right)$, $Q_{t_{i},B}^{md}\left(A,u\right)$
and $Q_{t_{i},B}^{dw}\left(A,u\right)$ where $u$ is $r$ in the
BS HW model and $v$ in the Heston model. These polynomials are supposed
to have all the same degree $d$. Here the most important remarks
\begin{itemize}
\item Create a grid of constant points $\mathcal{G}=\mathcal{A}\times\mathcal{B}=\left\{ \left(a_{k},b_{l}\right),0\leq k\leq K,0\leq l\leq L\right\} $
to be used as initial values to diffuse the couple $\left(A,B\right)$
using random scenarios. This grid lets us be sure that at each event
time, the set of initial values is well distributed and useful for
polynomial regression. In our tests we used $\mathcal{B}$ as set
of uniform nodes from $0$ to $P$ with $L=P/G$ : $\mathcal{B}=\left\{ 0,G,2G,\dots,P\right\} $.
The set $\mathcal{A}$ is more complicated. It contains points from
$A_{min}=0$ to $A_{max}=3P$; we also tried other values for $A_{max}$,
and $3P$ gave the best results. For each level of $B\in\mathcal{B}$,
we divide the interval $\left[0,3P\right]$ in 3 subsets: $DW_{B}=\left[0,\frac{1}{2}B\right]$,
$MD_{B}=\left[\frac{1}{2}B,\frac{3}{2}B\right]$ and $UP_{B}=\left[\frac{3}{2}B,3P\right]$.
In each of these subsets we define $d+1$ Chebychev nodes. These nodes
defines the grid. See Figure \ref{fig:The-grids}.
\item For each level $B$, the polynomials $Q_{t_{i},B}^{up}\left(A,u\right)$,
$Q_{t_{i},B}^{md}\left(A,u\right)$ and $Q_{t_{i},B}^{dw}\left(A,u\right)$
are obtained by regression, diffusing the state parameters of the
policy from the nodes in the sets $DW_{B}$, $MD_{B}$ and $UP_{B}$.
\item Use shift and scaling technique to improve regression.
\item As remarked before, the optimal withdrawal at last event time $t_{i}=T_{2}$,
is always equal to $\min\left(G,B_{T_{2}}^{-}\right)$.
\item To find the best value for the withdrawal amount $W_{t_{i}}$, numerical
tests proved that, if $G$ divides exactly $P$, then it's enough
to search among the multiples of $G$. This means that when we search
the best withdrawals, the possible value of $B$ are those of $\mathcal{B}$.
\end{itemize}
Here a pseudo code:

\begin{Verbatim}[frame=single,fontsize=\footnotesize,numbers=left]
Regression_by_lines(){
	int ETs= T2*WD_rate;	int H=P/G;
	Scenario_generation_step();
	for(int ti= ETs-1;ti>0;ti--){
		for(int l=0;l<L+1;l++){
			for(sector_l= DW_l, MD_l, UP_l){
				if(sector_l is not empty){			
				  Forward_Dynamic_step(ti,l,sector_l);
 			 	Backward_step_GMWB(ti);
 			 	Least_Squares_step(ti,l,sector_l);
	}}}}	 
	Last_Forward_Dynamic_step();
	Backward_step(0); 
}
\end{Verbatim}

The functions that we used are the following ones:
\begin{itemize}
\item Scenario\_generation\_step(). Generate the scenarios: $S$ and $v$
or $r$ .
\item Forward\_Dynamic\_step(ti,l,DW). For all the scenarios, setting $B_{t_{i}}=b_{l}=l\cdot P/G$,
and choosing $A_{t_{i}}$ in the node set $DW_{B_{t_{i}}}$, calculate
the state parameters of the policy at all the event times $t_{j}$
after $t_{i}$, using at each time $t_{j}$ the best withdrawal. This
functions does the same for the the sectors $MD_{B_{t_{i}}}$ and
$UP_{B_{t_{i}}}$.
\item Backward\_step\_GMWB(ti). For all the scenarios, calculate the value
of the policy $V_{t_{i}}^{s}$ at the event time $t_{i}$ as the sum
of discounted future cash flows.
\item Least\_Squares\_step(ti,l,DW). Perform polynomial regression. Calculate
$Q_{t_{i},B}^{dw}\left(A,u\right)$ using the value $\left(A_{t_{i}},B_{t_{i}},u_{t_{i}},V_{t_{i}}\right)$
diffused. This functions does the same for the the sectors $MD_{B_{t_{i}}}$
and $UP_{B_{t_{i}}}$.
\item Last\_Forward\_Dynamic\_step(). For all the scenarios, compute the
state parameters of the policy, starting from $t=0$ and $A_{0}=B_{0}=P$.
\end{itemize}

\paragraph*{Optimal surrender}

This case concerns GMWB-YD products. In their articles, Yang and Dai
suppose the PH to be entitled to surrender when optimal.

In this case we suppose that at each event time $t_{i}\in\left\{ t_{1},\dots,t_{N}\right\} $
the PH can withdraw the contract amount, or fully surrender. As we
did before, similarity reduction let us fix the value of $G$. We
denote $\mathcal{V}\left(A,t\right)$ the expected value at time $t$
of a generic policy whose state parameter is $A$ (similarity reduction
let us use only $A$ as variable) :
\[
\mathcal{V}\left(A,t\right)=\mathbb{E}\left[V_{s}\left(A,t\right)\right].
\]
So, we suppose that the PH surrenders at time $t_{i^{*}}$ if 
\[
\left(1-\kappa\right)\max\left(A_{t_{i^{*}}^{-}}-G,0\right)\geq\mathcal{V}\left(\max\left(A_{t_{i^{*}}^{-}}-G,0\right),t\right).
\]
The expected value $\mathcal{V}$ can be calculated with a standard
Longstaff-Schwartz approach:
\begin{enumerate}
\item Simulate $N$ random scenarios and price the policy into these scenarios
assuming that the PH follows a static approach.
\item For $t_{i}=t_{N}$ to $t_{i}=t_{1}$:

\begin{enumerate}
\item Approximate the function $\mathcal{V}\left(A,t_{i}\right)$ using
the least squares projection into a space of functions (usually polynomials).
\item For each scenario evaluate if $t_{i}$ is the good stopping time.
\end{enumerate}
\item Use at time $T_{1}$ similarity reduction to include account value
reset.
\item Calculate the average of the initial value $V_{s}\left(P,0\right)$
for all the scenarios to approximate $\mathcal{V}\left(P,0\right)$.
\end{enumerate}

\subsection{Standard Monte Carlo Method \label{MC}}

The Monte Carlo method is very similar to the Hybrid Monte Carlo one.
The only difference is the way we produce the random scenarios. The
projection phase is the same as Hybrid Monte Carlo one.

\subsubsection{Scenario generation}

We distinguish two cases for the two models.

\paragraph*{The Heston Model}

The generation of the scenarios (underlying and volatility) in this
case has been done using a third order scheme described in Alfonsi
\cite{AA}.

\paragraph*{The Black-Scholes Hull-White Model}

The generation of the scenarios (underlying and interest rate) in
this case has been done using an exact scheme described in Ostrovski
\cite{VO}, with a few changes in order to incorporate the correlation
between underlying and interest rate.

\subsection{PDE Hybrid Method}

The Hybrid PDE approach is different from the previous ones. In fact
it's a PDE pricing method and it's based on Briani et al. \cite{BCZ0},
\cite{BCZ} both for Heston and Hull-White case. Using a tree to diffuse
volatility or interest rate, we freeze these values between two tree-levels
and we solve a Black Scholes PDE for each node of the tree, using
as initial data a weighted mix of the data of the upcoming nodes.

We can resume the pricing methods in three features: model, algorithm
structure and pricing.

We start describing the model between the event times.

\subsubsection{The Heston Model }

Starting from the model for the found $S_{t}$ in (\ref{eq:He}),
we call $\bar{\rho}=\sqrt{1-\rho^{2}}$ and we write $Z_{t}^{A}=\rho Z_{t}^{v}+\bar{\rho}\bar{Z}_{t}^{A}$,
where $\bar{Z}^{A}$ is a Brownian motion uncorrelated with $Z^{v}$.
Then,
\[
\begin{cases}
dA_{t}=\left(r-\alpha_{tot}\right)A_{t}dt+\sqrt{v_{t}}A_{t}\left(\rho dZ_{t}^{v}+\bar{\rho}d\bar{Z}_{t}^{A}\right) & v_{0}=\bar{v}_{0},\\
dv_{t}=k\left(\theta-v_{t}\right)dt+\omega\sqrt{v_{t}}dZ_{t}^{V} & A_{0}=\bar{A}_{0},
\end{cases}\ \ \ d\left\langle \bar{Z}_{t}^{A},Z_{t}^{v}\right\rangle =0,
\]
covering the behavior of $A_{t}$ between two event times, we define
the process 
\[
Y_{t}^{E}=\ln\left(A_{t}\right)-\frac{\rho}{\omega}v_{t},\ Y_{0}^{E}=\ln\left(A_{0}\right)-\frac{\rho}{\omega}v_{0}
\]

Then, 
\begin{equation}
A_{t}=\exp\left(Y_{t}^{E}+\frac{\rho}{\omega}v_{t}\right)\label{eq:inv_He}
\end{equation}

and 
\[
dY_{t}^{E}=\left(r-\alpha_{tot}-\frac{v_{t}}{2}-\frac{\rho}{\omega}k\left(\theta-v_{t}\right)\right)dt+\bar{\rho}\sqrt{v_{t}}d\bar{Z}_{t}^{A}.
\]

This process $Y^{E}$ is important because it's a process uncorrelated
with the volatility process $v$, and we introduce it as in \cite{BCZ0}.
We are going to use it to define a PDE to be solved along the tree.

We define $\hat{\mathcal{V}}^{He}\left(t,Y_{t}^{E}\right)=\mathcal{V}\left(t,A_{t}\right)$. 

If, in a small time lag $\left[\tau,\tau+\Delta\tau\right]$, we approximate
the process $Y_{t}^{E}$ by the process $\bar{Y}_{t}^{E}$ whose dynamics
is given by 
\[
d\bar{Y}_{t}^{E}=\left(r-\alpha_{tot}-\frac{v_{\tau}}{2}-\frac{\rho}{\omega}k\left(\theta-v_{\tau}\right)\right)dt+\bar{\rho}\sqrt{v_{\tau}}d\bar{Z}_{t}^{A}.
\]

Then, $\hat{\mathcal{V}}^{He}\left(t,\bar{Y}_{t}^{E}\right)$ verifies
the following PDE
\begin{equation}
\frac{\partial\hat{\mathcal{V}}^{He}}{\partial t}+\left(r-\alpha_{tot}-\frac{v_{\tau}}{2}-\frac{\rho}{\omega}k\left(\theta-v_{\tau}\right)\right)\frac{\partial\hat{\mathcal{V}}^{He}}{\partial\bar{Y}_{t}^{E}}+\frac{\bar{\rho}^{2}v_{\tau}}{2}\frac{\partial^{2}\hat{\mathcal{V}}^{He}}{\partial^{2}\bar{Y}_{t}^{E}}-r\hat{\mathcal{V}}^{He}=0.\label{eq:PDE-He}
\end{equation}

\subsubsection{The Black-Scholes Hull-White Model}

Starting from the model for the found $S_{t}$ in (\ref{eq:HW}),
we call $\bar{\rho}=\sqrt{1-\rho^{2}}$ and we write $Z_{t}^{A}=\rho Z_{t}^{r}+\bar{\rho}\bar{Z}_{t}^{A}$,
where $\bar{Z}^{A}$ is a Brownian motion uncorrelated with $Z^{r}$.
Then,

\[
\begin{cases}
dA_{t}=A_{t}\left(r-\alpha_{tot}\right)dt+\sigma A_{t}\left(\rho dZ_{t}^{r}+\bar{\rho}d\bar{Z}_{t}^{A}\right) & A_{0}=\bar{A}_{0},\\
dX_{t}=-kX_{t}dt+dZ_{t}^{r} & X_{0}=0,\\
r_{t}=\omega X_{t}+\beta\left(t\right),
\end{cases}\ \ \ d\left\langle \bar{Z}_{t}^{A},Z_{t}^{r}\right\rangle =0.
\]
We define the process 
\[
Y_{t}^{U}=\ln\left(A_{t}\right)-\rho\sigma X_{t},\ Y_{0}^{U}=\ln\left(A_{0}\right)
\]

Then, 
\begin{equation}
A_{t}=\exp\left(Y_{t}+\rho\sigma X_{t}\right)\label{eq:inv_HW}
\end{equation}

and 
\[
dY_{t}^{U}=\left(r_{t}-\alpha_{tot}-\frac{\sigma^{2}}{2}+\sigma\rho kX_{t}\right)dt+\sigma\bar{\rho}d\bar{Z}_{t}^{A}.
\]

This process $Y^{U}$ is important because it's a process uncorrelated
with the mean-reverting process $X$, and we introduce it as in \cite{BCZ0}.
We are going to use it to define a PDE to be solved along the tree.

We define $\hat{\mathcal{V}}^{HW}\left(t,Y_{t}^{U}\right)=\mathcal{V}\left(t,A_{t}\right)$.
If, in a small time lag $\left[\tau,\tau+\Delta\tau\right]$, we approximate
the process $Y_{t}^{U}$ by the process $\bar{Y}_{t}^{U}$ whose dynamics
is given by
\[
d\bar{Y}_{t}^{U}=\left(r_{\tau}-\alpha_{tot}-\frac{\sigma^{2}}{2}+\sigma\rho kX_{\tau}\right)dt+\sigma\bar{\rho}dZ_{t},
\]

and the interest rate process by $r_{\tau}$, then, $\hat{\mathcal{V}}^{HW}\left(t,\bar{Y}_{t}^{U}\right)$
verifies the following PDE
\begin{equation}
\frac{\partial\hat{\mathcal{V}}^{HW}}{\partial t}+\left(r_{\tau}-\alpha_{tot}-\frac{\sigma^{2}}{2}+\sigma\rho kX_{\tau}\right)\frac{\partial\hat{\mathcal{V}}^{HW}}{\partial\bar{Y}_{t}^{U}}+\frac{\bar{\rho}^{2}\sigma^{2}}{2}\frac{\partial^{2}\hat{\mathcal{V}}^{HW}}{\partial^{2}\bar{Y}_{t}^{U}}-r_{\tau}\hat{\mathcal{V}}^{HW}=0.\label{eq:PDE-HW}
\end{equation}

\subsubsection{Algorithm structure}

The structures for this algorithm consist in a tree and a PDE solver.
As described in Briani et al. \cite{BCZ0}, \cite{BCZ}, we use a
tree to diffuse the volatility (or the interest rate) along the life
of the product, and we solve backward a 1D PDE freezing at each node
of the tree the volatility (or the interest rate). The tree is built
according to Section \ref{tree} (quadrinomial tree, matching the
first three moments of the process), and the PDE is solved with a
finite difference approach. We have to solve the PDE between event
times, and at each event time we apply the changes to the states to
reproduce the effects of the events.

We remark that we solve the PDE doing a single time step that requires
only a linear complexity because we have to solve a linear system
with tridiagonal matrix. The computational cost is low as observed
in \cite{BCZ0} and \cite{BCZ}. We observe that $X_{t}$ and $V_{t}$
processes are mean reverting. Thanks to the way the trees are built,
there are many nodes in the trees that cannot be visited by the approximating
Markov chain. Therefore their probability $p_{n,j}$ to be visited
is worth $0$ and they have no impact on the values at the root of
the tree. There is no reason to do any operation for those nodes.
So, to save time, we do the standard step (mix up the vectors according
to the transition probabilities and solve backward a PDE) only for
those nodes having $p_{n,j}>0$. This curtailing technique reduces
the computational time, and the convergence of the method is preserved.
A similar approach is used in \cite{AW}.

\subsubsection{\label{Pric}Pricing}

We distinguish 3 cases.

\subsubsection*{Static case}

This case is common to both GMWB-CF and GMWB-YD products. The problem
dimension is 2: about GMWB-CF, at each event time the value of the
the base benefit $B_{t_{i}}$ is equal to $P-G\cdot i$ and thus it's
not a problem variable; about GMWB-YD similarity reduction reduce
problem dimension to 2.

For each node of the tree we have to solve one PDE using the mixture
of the the data of the upcoming nodes: the mixture is done according
to transition probabilities. The PDE to be solved are those in (\ref{eq:PDE-He})
and (\ref{eq:PDE-HW}) where $\left[\tau,\tau+\Delta\tau\right]$
denotes the time lag between two tree nodes.

The variables $\bar{r}$, $\bar{X}$ and $\bar{v}$ will denote the
frozen values of $r_{t}$, $X_{t}$ and $v_{t}$ using the data of
the actual node. We used a finite differences approach using equally
spaced nodes for $Y_{t}$ processes. To reduce the run time, we do
this only for most relevant nodes: this cutting technique dramatically
reduced calculation times without compromising the quality of results.
Then, using the inverse transformations (\ref{eq:inv_He}) and (\ref{eq:inv_HW}),
we apply the event times actions in (\ref{eq:ET-CF}) or equivalently
(\ref{eq:ET-YD}).

\subsubsection*{Optimal Withdrawal}

This case is about GMWB-CF products. This is the hardest to be treated
because the problem dimension is 3. We solve the same PDE as in Static
case, but this time we have to solve them for different values of
$B_{t}$ and chose the best withdrawal $W_{t}$ at each event time.
Numerical test showed that it's enough to search the best withdrawal
between multiples of $G$ equal or smaller than the base benefit.
Then we decided to solve the problem for all $B_{t}$ values of which
are multiples of $G$ and are smaller than the initial premium $P$:
$B=0$, $B=G$, $B_{t}=2G$, $\dots$$B=nG=P$. Then, we solve $n$
2-dimensional problems rather then one 3-dimensional problem. This
approach is similar to ``Regression by Lines'' defined for MC methods.

Best withdrawal search is performed searching among permissible withdrawals
which are multiples of $G$: $W=0,$ $W=G,$$\dots$$W=mG=B$. The
estimate of $\mathcal{V}\left(A,B\right)$ for those values of $A$
that aren't on the grid, is done using splines. 

\begin{figure}
\begin{centering}
\includegraphics[scale=0.9]{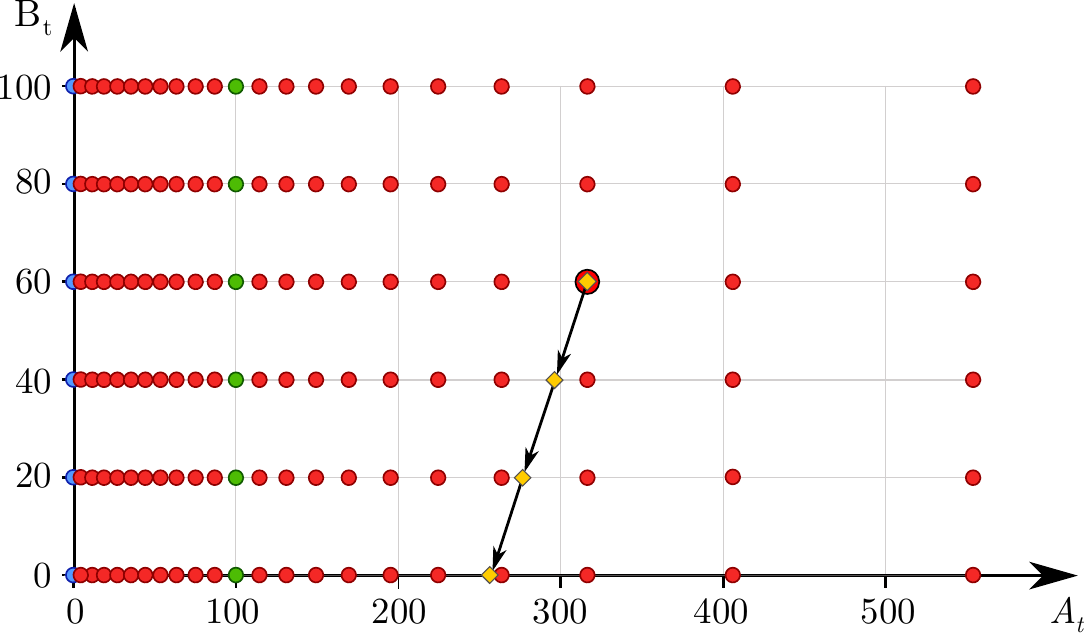}
\par\end{centering}

\caption{\label{fig:Optimal-CF}The scheme for PDE optimal withdrawal for a
given node.}

\end{figure}

In Figure \ref{fig:Optimal-CF} we can see a scheme that represents
what happens for a product with $G=20$: for example, a $5$ years
maturity GM with annual withdrawal rate and $P=100$ ($G=20$). Nodes
are exponentially distributed (uniformly for $Y$ process) and for
each $B$ value, we add a node that represents $A=0$ (blue nodes).
For each node, first we mix the data vectors of the upcoming nodes
according to transition probabilities. Then we solve a PDE backward
starting from the mixture of the data. Then we apply withdrawal step:
for each node we consider admissible withdrawals of the type $W=kG$
and we chose the value that maximize PH's benefit: cash flow plus
policy value. This research is shown in the Figure (see yellow nodes
that corresponds to possible withdrawals).

\subsubsection*{Optimal surrender}

This case is about GMWB-YD products. It's much more simpler than optimal
withdrawal. In fact, the PH can only chose between withdrawal at the
contract rate and fully surrender.

Withdrawal step at event time $t_{i}$ consists into replacing $\mathcal{V}\left(A,t_{i}\right)$
by
\[
\max\left[G+\mathcal{V}\left(\max\left(A-G,0\right),t_{i}\right);G+\left(1-\kappa\right)\left(A-G\right)\right].
\]

\subsection{PDE ADI Method}

We propose a PDE pricing method with alternative direction implicit
scheme which has been already successfully used for European financial
product (see \cite{HH}) and for an insurance GLWB product (see \cite{GMZ}).
This method permits to treat the Heston model and the BS HW model.
This method is fast and accurate. Moreover it is easy to take into
account the similarity reduction and the optimal behavior. For this
method, we followed the same principles of HPDE method about taking
in account the event times.

The PDEs to be solved in the two models are 

{\small{} \[ \mathcal{V}_{t}^{He}+\frac{VA^{2}}{2}\mathcal{V}_{AA}^{He} + \frac{\omega^{2}V}{2}\mathcal{V}_{VV}^{He} +\left(r-\alpha_{tot}\right)A\mathcal{V}_{A}^{He}+\rho \omega AV \mathcal{V}_{AV}^{He}+k \left(\theta -V\right)\mathcal{V}_{V}^{He}-r\mathcal{V}^{He}=0\ \tag{He} \] }{\small \par}

{\small{} \[ \mathcal{V}_{t}^{HW}+\frac{\sigma^{2}A^{2}}{2}\mathcal{V}_{AA}^{HW} + \frac{\omega^{2}}{2}\mathcal{V}_{rr}^{HW} +\left(r-\alpha_{tot}\right)A\mathcal{V}_{A}^{HW}+\rho \omega A\sigma \mathcal{V}_{Ar}^{HW}+k \left(\theta_{t} -r\right)\mathcal{V}_{r}^{HW}-r\mathcal{V}^{HW}=0\ \tag{HW} \] }{\small \par}

There are multiple numerical parameters which have to be carefully
chosen. We have to choose the grids for the benefit base, the account
value, the rate in the BS HW model and the volatility in the Heston
model. We have chosen to use the  meshes described in \cite{HH} with the parameters \[ A_{left} = 0.8 S_{0}\quad A_{right}=1.2S_{0} \quad A_{max}=1000\cdot T2\cdot S_{0}  \quad \text{ and } d_{1} = S_{0}/20, \] for the mesh of variable $A$,  \[ R_{\max} = 0.8, \quad c=R_{0}\quad \text{ and } d_{2} = R_{\max}/400 \] for the mesh of variable $r$ in the BS HW model, and \[ V_{\max} = MIN(MAX(100V_{0},1),5) \quad \text{ and } d_{3} = V_{\max}/500. \] for the mesh of variable $V$ in the Heston model. Some
grid are uniform or based on hyperbolic transformation. Moreover the
boundary conditions are completely unknown, and an asymptotic study
would be necessary to chose them. We have chosen homogeneous Neumann
boundary conditions, and we have chosen very large grids to avoid
that this choice impacts the results. We have only used the Douglas
scheme, but other schemes are possible to have better order of convergence
in time. Thus many possibilities are possible to improve the ADI scheme,
but the easier is already enough to obtain good results.

\section{\label{5}Numerical Results}

In this Section we compare the numerical methods used in Section \ref{4}:
Hybrid Monte Carlo (\emph{HMC}), Standard Monte Carlo (\emph{SMC}),
Hybrid PDE (\emph{HPDE}), and ADI PDE (\emph{APDE}). In particular
we compare pricing and Greeks computation in \emph{Static Case} and
\emph{Dynamic Case} for both the two product types.

We chose the parameters of the methods according to 4 configurations
(\emph{A, B, C, D}), with an increasing number of steps and so that
the calculation time for the various methods in each configuration
were close. The 4 configurations are in Table \ref{tab:Configuration-parameters}
and in Table \ref{tab:Configuration-parameters-1} with the notation
(time steps per year $\times$ space steps $\times$ vol steps) for
the ADI PDE method, (time steps per year $\times$ space steps ) for
the Hybrid PDE method and (time steps per year $\times$ number of
simulations) for the MC ones. In Monte Carlo for Dynamic case, we
also add the degree of the approximating polynomial. These values
have been chosen to achieve approximately these run times using an
ordinary computer: $\left(A\right)$ 30 s, $\left(B\right)$ 120 s,
$\left(C\right)$ 480 s, $\left(D\right)$ 1920 s. To reduce the run
time we do the secant iterations using an increasing number of time
steps for all the methods: the values in Table \ref{tab:Configuration-parameters}
and \ref{tab:Configuration-parameters-1} are those used for the last
3 iterations.

We use the Standard MC both as a pricing method, both as a benchmark
(BM). About the benchmark, in the Static case we used $10^{8}$ independent
runs. In the Dynamic case we used $10^{6}$ independent runs; in each
sub runs the expected value has been approximated by a $4^{th}$ degree
polynomial. 

The search for the fair $\alpha_{g}$ value has been driven by the
secant method. The initial values for this method were $\alpha_{g}=0$
bp and $\alpha_{g}=200$ bp.

To achieve Delta calculation in Monte Carlo methods we used a $1\permil$
shock in Static case and $1\%$ in Dynamic case.

\begin{table}
\begin{centering}
{\footnotesize{}}%
\begin{tabular}{|>{\centering}p{0.15cm}|>{\centering}p{1.6cm}>{\centering}p{1.6cm}>{\centering}p{1.4cm}>{\centering}p{1.7cm}||>{\centering}p{1.6cm}>{\centering}p{1.6cm}>{\centering}p{1.4cm}>{\centering}p{1.7cm}|}
\cline{2-9} 
\multicolumn{1}{>{\centering}p{0.15cm}|}{} & \multicolumn{4}{c||}{\textsc{\small{}BS HW Static}} & \multicolumn{4}{c|}{\textsc{\small{}Heston Static}}\tabularnewline
\multicolumn{1}{>{\centering}p{0.15cm}|}{} & \textsc{\footnotesize{}HMC} & \textsc{\footnotesize{}SMC} & \textsc{\footnotesize{}HPDE} & \textsc{\footnotesize{}APDE} & \textsc{\footnotesize{}HMC} & \textsc{\footnotesize{}SMC} & \textsc{\footnotesize{}HPDE} & \textsc{\footnotesize{}APDE}\tabularnewline
\hline 
\emph{\footnotesize{}A} & \textcolor{blue}{\scriptsize{}$4\negthinspace\times\negthinspace9.2\negthinspace\cdot\negthinspace10^{5}$} & \textcolor{blue}{\scriptsize{}$1\negthinspace\times\negthinspace1.7\negthinspace\cdot\negthinspace10^{6}$} & \textcolor{blue}{\scriptsize{}$260\negthinspace\times\negthinspace250$} & \textcolor{blue}{\scriptsize{}$25\negthinspace\times\negthinspace250\!\times505$} & \textcolor{blue}{\scriptsize{}$4\negthinspace\times\negthinspace5.8\negthinspace\cdot\negthinspace10^{5}$} & \textcolor{blue}{\scriptsize{}$4\negthinspace\times\negthinspace5.2\negthinspace\cdot\negthinspace10^{5}$} & \textcolor{blue}{\scriptsize{}$270\negthinspace\times\negthinspace250$} & \textcolor{blue}{\scriptsize{}$25\negthinspace\times\negthinspace250\!\times505$}\tabularnewline
\emph{\footnotesize{}B} & \textcolor{blue}{\scriptsize{}$8\negthinspace\times\negthinspace1.8\negthinspace\cdot\negthinspace10^{6}$} & \textcolor{blue}{\scriptsize{}$1\negthinspace\times\negthinspace5.7\negthinspace\cdot\negthinspace10^{6}$} & \textcolor{blue}{\scriptsize{}$420\negthinspace\times\negthinspace500$} & \textcolor{blue}{\scriptsize{}$40\negthinspace\times\negthinspace400\!\times85$} & \textcolor{blue}{\scriptsize{}$8\negthinspace\times\negthinspace1.2\negthinspace\cdot\negthinspace10^{6}$} & \textcolor{blue}{\scriptsize{}$8\negthinspace\times\negthinspace1.2\negthinspace\cdot\negthinspace10^{6}$} & \textcolor{blue}{\scriptsize{}$520\negthinspace\times\negthinspace500$} & \textcolor{blue}{\scriptsize{}$40\negthinspace\times\negthinspace400\!\times80$}\tabularnewline
\emph{\footnotesize{}C} & \textcolor{blue}{\scriptsize{}$12\negthinspace\times\negthinspace6.3\negthinspace\cdot\negthinspace10^{6}$} & \textcolor{blue}{\scriptsize{}$1\negthinspace\times\negthinspace2.9\negthinspace\cdot\negthinspace10^{7}$} & \textcolor{blue}{\scriptsize{}$780\negthinspace\times\negthinspace1000$} & \textcolor{blue}{\scriptsize{}$60\negthinspace\times\negthinspace620\!\times125$} & \textcolor{blue}{\scriptsize{}$12\negthinspace\times\negthinspace3.9\negthinspace\cdot\negthinspace10^{6}$} & \textcolor{blue}{\scriptsize{}$12\negthinspace\times\negthinspace3.4\negthinspace\cdot\negthinspace10^{6}$} & \textcolor{blue}{\scriptsize{}$850\negthinspace\times\negthinspace1000$} & \textcolor{blue}{\scriptsize{}$60\negthinspace\times\negthinspace620\!\times120$}\tabularnewline
\emph{\footnotesize{}D} & \textcolor{blue}{\scriptsize{}$16\negthinspace\times\negthinspace1.9\negthinspace\cdot\negthinspace10^{7}$} & \textcolor{blue}{\scriptsize{}$1\negthinspace\times\negthinspace1.2\negthinspace\cdot\negthinspace10^{8}$} & \textcolor{blue}{\scriptsize{}$1200\negthinspace\times\negthinspace2000$} & \textcolor{blue}{\scriptsize{}$100\negthinspace\times\negthinspace1000\!\times215$} & \textcolor{blue}{\scriptsize{}$16\negthinspace\times\negthinspace1.2\negthinspace\cdot\negthinspace10^{7}$} & \textcolor{blue}{\scriptsize{}$16\negthinspace\times\negthinspace1.1\negthinspace\cdot\negthinspace10^{7}$} & \textcolor{blue}{\scriptsize{}$1400\negthinspace\times\negthinspace2000$} & \textcolor{blue}{\scriptsize{}$100\negthinspace\times\negthinspace1000\!\times200$}\tabularnewline
\hline 
\end{tabular}
\par\end{centering}{\footnotesize \par}

{\footnotesize{}\vspace{0.1cm}}{\footnotesize \par}

\begin{centering}
{\footnotesize{}}%
\begin{tabular}{|>{\centering}p{0.15cm}|>{\centering}p{1.6cm}>{\centering}p{1.6cm}>{\centering}p{1.4cm}>{\centering}p{1.6cm}||>{\centering}p{1.6cm}>{\centering}p{1.6cm}>{\centering}p{1.4cm}>{\centering}p{1.7cm}|}
\cline{2-9} 
\multicolumn{1}{>{\centering}p{0.15cm}|}{} & \multicolumn{4}{c||}{\textsc{\small{}BS HW Dynamic}} & \multicolumn{4}{c|}{\textsc{\small{}Heston Dynamic}}\tabularnewline
\multicolumn{1}{>{\centering}p{0.15cm}|}{} & \textsc{\footnotesize{}HMC} & \textsc{\footnotesize{}SMC} & \textsc{\footnotesize{}HPDE} & \textsc{\footnotesize{}APDE} & \textsc{\footnotesize{}HMC} & \textsc{\footnotesize{}SMC} & \textsc{\footnotesize{}HPDE} & \textsc{\footnotesize{}APDE}\tabularnewline
\hline 
\emph{\footnotesize{}A} & \textcolor{blue}{\scriptsize{}$4\negthinspace\times\negthinspace6.0\!\cdot\!10^{4}\negthinspace\times\negthinspace1$} & \textcolor{blue}{\scriptsize{}$1\negthinspace\times\negthinspace6.5\!\cdot\!10^{4}\negthinspace\times\negthinspace1$} & \textcolor{blue}{\scriptsize{}$70\negthinspace\times\negthinspace250$} & \textcolor{blue}{\scriptsize{}$8\negthinspace\times\negthinspace95\!\times30$} & \textcolor{blue}{\scriptsize{}$4\negthinspace\times\negthinspace5.1\!\cdot\!10^{4}\negthinspace\times\negthinspace1$} & \textcolor{blue}{\scriptsize{}$4\negthinspace\times\negthinspace5.5\!\cdot\!10^{4}\negthinspace\times\negthinspace1$} & \textcolor{blue}{\scriptsize{}$88\negthinspace\times\negthinspace250$} & \textcolor{blue}{\scriptsize{}$10\negthinspace\times\negthinspace125\!\times25$}\tabularnewline
\emph{\footnotesize{}B} & \textcolor{blue}{\scriptsize{}$8\negthinspace\times\negthinspace8.7\!\cdot\!10^{4}\negthinspace\times\negthinspace2$} & \textcolor{blue}{\scriptsize{}$1\negthinspace\times\negthinspace9.5\!\cdot\!10^{4}\negthinspace\times\negthinspace2$} & \textcolor{blue}{\scriptsize{}$160\negthinspace\times\negthinspace500$} & \textcolor{blue}{\scriptsize{}$14\negthinspace\times\negthinspace150\!\times48$} & \textcolor{blue}{\scriptsize{}$8\negthinspace\times\negthinspace1.2\!\cdot\!10^{4}\negthinspace\times\negthinspace2$} & \textcolor{blue}{\scriptsize{}$8\negthinspace\times\negthinspace1.3\!\cdot\!10^{4}\negthinspace\times\negthinspace2$} & \textcolor{blue}{\scriptsize{}$160\negthinspace\times\negthinspace500$} & \textcolor{blue}{\scriptsize{}$15\negthinspace\times\negthinspace200\!\times40$}\tabularnewline
\emph{\footnotesize{}C} & \textcolor{blue}{\scriptsize{}$12\negthinspace\times\negthinspace1.8\!\cdot\!10^{5}\negthinspace\times\negthinspace3$} & \textcolor{blue}{\scriptsize{}$1\negthinspace\times\negthinspace1.9\!\cdot\!10^{5}\negthinspace\times\negthinspace3$} & \textcolor{blue}{\scriptsize{}$270\negthinspace\times\negthinspace1000$} & \textcolor{blue}{\scriptsize{}$22\negthinspace\times\negthinspace250\!\times75$} & \textcolor{blue}{\scriptsize{}$12\negthinspace\times\negthinspace2.3\!\cdot\!10^{5}\negthinspace\times\negthinspace3$} & \textcolor{blue}{\scriptsize{}$12\negthinspace\times\negthinspace2.5\!\cdot\!10^{5}\negthinspace\times\negthinspace3$} & \textcolor{blue}{\scriptsize{}$266\negthinspace\times\negthinspace1000$} & \textcolor{blue}{\scriptsize{}$25\negthinspace\times\negthinspace320\!\times60$}\tabularnewline
\emph{\footnotesize{}D} & \textcolor{blue}{\scriptsize{}$16\negthinspace\times\negthinspace3.5\!\cdot\!10^{5}\negthinspace\times\negthinspace4$} & \textcolor{blue}{\scriptsize{}$1\negthinspace\times\negthinspace3.5\!\cdot\!10^{5}\negthinspace\times\negthinspace4$} & \textcolor{blue}{\scriptsize{}$360\negthinspace\times\negthinspace2000$} & \textcolor{blue}{\scriptsize{}$35\negthinspace\times\negthinspace400\!\times120$} & \textcolor{blue}{\scriptsize{}$16\negthinspace\times\negthinspace4.2\!\cdot\!10^{5}\negthinspace\times\negthinspace4$} & \textcolor{blue}{\scriptsize{}$16\negthinspace\times\negthinspace5.0\!\cdot\!10^{5}\negthinspace\times\negthinspace4$} & \textcolor{blue}{\scriptsize{}$350\negthinspace\times\negthinspace2000$} & \textcolor{blue}{\scriptsize{}$40\negthinspace\times\negthinspace500\!\times90$}\tabularnewline
\hline 
\end{tabular}
\par\end{centering}{\footnotesize \par}

\caption{\label{tab:Configuration-parameters} Configuration parameters for
the BS HW model and for the Heston model, Static and Dynamic for the
GMWB-CF product with $T_{2}=10$ and $WF=1$.}
\end{table}

\begin{table}
\begin{centering}
{\footnotesize{}}%
\begin{tabular}{|>{\centering}p{0.15cm}|>{\centering}p{1.7cm}>{\centering}p{1.7cm}>{\centering}p{1.4cm}>{\centering}p{1.6cm}||>{\centering}p{1.7cm}>{\centering}p{1.7cm}>{\centering}p{1.4cm}>{\centering}p{1.8cm}|}
\cline{2-9} 
\multicolumn{1}{>{\centering}p{0.15cm}|}{} & \multicolumn{4}{c||}{\textsc{\small{}BS HW Static}} & \multicolumn{4}{c|}{\textsc{\small{}Heston Static}}\tabularnewline
\multicolumn{1}{>{\centering}p{0.15cm}|}{} & \textsc{\footnotesize{}HMC} & \textsc{\footnotesize{}SMC} & \textsc{\footnotesize{}HPDE} & \textsc{\footnotesize{}APDE} & \textsc{\footnotesize{}HMC} & \textsc{\footnotesize{}SMC} & \textsc{\footnotesize{}HPDE} & \textsc{\footnotesize{}APDE}\tabularnewline
\hline 
\emph{\footnotesize{}A} & \textcolor{blue}{\scriptsize{}$4\negthinspace\times\negthinspace3.2\negthinspace\cdot\negthinspace10^{5}$} & \textcolor{blue}{\scriptsize{}$1\negthinspace\times\negthinspace6.0\negthinspace\cdot\negthinspace10^{5}$} & \textcolor{blue}{\scriptsize{}$130\negthinspace\times\negthinspace250$} & \textcolor{blue}{\scriptsize{}$10\negthinspace\times\negthinspace245\!\times50$} & \textcolor{blue}{\scriptsize{}$4\negthinspace\times\negthinspace2.3\negthinspace\cdot\negthinspace10^{5}$} & \textcolor{blue}{\scriptsize{}$4\negthinspace\times\negthinspace2.0\negthinspace\cdot\negthinspace10^{5}$} & \textcolor{blue}{\scriptsize{}$120\negthinspace\times\negthinspace250$} & \textcolor{blue}{\scriptsize{}$10\negthinspace\times\negthinspace250\!\times50$}\tabularnewline
\emph{\footnotesize{}B} & \textcolor{blue}{\scriptsize{}$8\negthinspace\times\negthinspace6.4\negthinspace\cdot\negthinspace10^{5}$} & \textcolor{blue}{\scriptsize{}$1\negthinspace\times\negthinspace2.3\negthinspace\cdot\negthinspace10^{6}$} & \textcolor{blue}{\scriptsize{}$215\negthinspace\times\negthinspace500$} & \textcolor{blue}{\scriptsize{}$15\negthinspace\times\negthinspace375\!\times80$} & \textcolor{blue}{\scriptsize{}$8\negthinspace\times\negthinspace4.6\negthinspace\cdot\negthinspace10^{5}$} & \textcolor{blue}{\scriptsize{}$8\negthinspace\times\negthinspace3.8\negthinspace\cdot\negthinspace10^{5}$} & \textcolor{blue}{\scriptsize{}$220\negthinspace\times\negthinspace500$} & \textcolor{blue}{\scriptsize{}$15\negthinspace\times\negthinspace380\!\times80$}\tabularnewline
\emph{\footnotesize{}C} & \textcolor{blue}{\scriptsize{}$12\negthinspace\times\negthinspace2.2\negthinspace\cdot\negthinspace10^{6}$} & \textcolor{blue}{\scriptsize{}$1\negthinspace\times\negthinspace1.2\negthinspace\cdot\negthinspace10^{7}$} & \textcolor{blue}{\scriptsize{}$415\negthinspace\times\negthinspace1000$} & \textcolor{blue}{\scriptsize{}$35\negthinspace\times\negthinspace520\!\times110$} & \textcolor{blue}{\scriptsize{}$12\negthinspace\times\negthinspace1.6\negthinspace\cdot\negthinspace10^{6}$} & \textcolor{blue}{\scriptsize{}$12\negthinspace\times\negthinspace1.3\negthinspace\cdot\negthinspace10^{6}$} & \textcolor{blue}{\scriptsize{}$425\negthinspace\times\negthinspace1000$} & \textcolor{blue}{\scriptsize{}$36\negthinspace\times\negthinspace530\!\times110$}\tabularnewline
\emph{\footnotesize{}D} & \textcolor{blue}{\scriptsize{}$16\negthinspace\times\negthinspace6.8\negthinspace\cdot\negthinspace10^{6}$} & \textcolor{blue}{\scriptsize{}$1\negthinspace\times\negthinspace4.5\negthinspace\cdot\negthinspace10^{7}$} & \textcolor{blue}{\scriptsize{}$480\negthinspace\times\negthinspace2000$} & \textcolor{blue}{\scriptsize{}$55\negthinspace\times\negthinspace880\!\times180$} & \textcolor{blue}{\scriptsize{}$16\negthinspace\times\negthinspace4.8\negthinspace\cdot\negthinspace10^{6}$} & \textcolor{blue}{\scriptsize{}$16\negthinspace\times\negthinspace4.0\negthinspace\cdot\negthinspace10^{6}$} & \textcolor{blue}{\scriptsize{}$480\negthinspace\times\negthinspace2000$} & \textcolor{blue}{\scriptsize{}$55\negthinspace\times\negthinspace890\!\times180$}\tabularnewline
\hline 
\end{tabular}
\par\end{centering}{\footnotesize \par}

{\footnotesize{}\vspace{0.1cm}}{\footnotesize \par}

\begin{centering}
{\footnotesize{}}%
\begin{tabular}{|>{\centering}p{0.15cm}|>{\centering}p{1.7cm}>{\centering}p{1.7cm}>{\centering}p{1.4cm}>{\centering}p{1.6cm}||>{\centering}p{1.7cm}>{\centering}p{1.7cm}>{\centering}p{1.4cm}>{\centering}p{1.8cm}|}
\cline{2-9} 
\multicolumn{1}{>{\centering}p{0.15cm}|}{} & \multicolumn{4}{c||}{\textsc{\small{}BS HW Dynamic}} & \multicolumn{4}{c|}{\textsc{\small{}Heston Dynamic}}\tabularnewline
\multicolumn{1}{>{\centering}p{0.15cm}|}{} & \textsc{\footnotesize{}HMC} & \textsc{\footnotesize{}SMC} & \textsc{\footnotesize{}HPDE} & \textsc{\footnotesize{}APDE} & \textsc{\footnotesize{}HMC} & \textsc{\footnotesize{}SMC} & \textsc{\footnotesize{}HPDE} & \textsc{\footnotesize{}APDE}\tabularnewline
\hline 
\emph{\footnotesize{}A} & \textcolor{blue}{\scriptsize{}$4\negthinspace\times\negthinspace6.8\negthinspace\cdot\negthinspace10^{4}\!\times1$} & \textcolor{blue}{\scriptsize{}$1\negthinspace\times\negthinspace8.1\negthinspace\cdot\negthinspace10^{4}\!\times1$} & \textcolor{blue}{\scriptsize{}$130\negthinspace\times\negthinspace250$} & \textcolor{blue}{\scriptsize{}$10\negthinspace\times\negthinspace245\!\times50$} & \textcolor{blue}{\scriptsize{}$4\negthinspace\times\negthinspace5.5\negthinspace\cdot\negthinspace10^{4}\!\times2$} & \textcolor{blue}{\scriptsize{}$4\negthinspace\times\negthinspace5.8\negthinspace\cdot\negthinspace10^{4}\!\times2$} & \textcolor{blue}{\scriptsize{}$120\negthinspace\times\negthinspace250$} & \textcolor{blue}{\scriptsize{}$10\negthinspace\times\negthinspace250\!\times50$}\tabularnewline
\emph{\footnotesize{}B} & \textcolor{blue}{\scriptsize{}$8\negthinspace\times\negthinspace2.5\negthinspace\cdot\negthinspace10^{5}\!\times2$} & \textcolor{blue}{\scriptsize{}$1\negthinspace\times\negthinspace3.4\negthinspace\cdot\negthinspace10^{5}\!\times2$} & \textcolor{blue}{\scriptsize{}$215\negthinspace\times\negthinspace500$} & \textcolor{blue}{\scriptsize{}$15\negthinspace\times\negthinspace375\!\times80$} & \textcolor{blue}{\scriptsize{}$8\negthinspace\times\negthinspace2.2\negthinspace\cdot\negthinspace10^{5}\!\times3$} & \textcolor{blue}{\scriptsize{}$8\negthinspace\times\negthinspace2.0\negthinspace\cdot\negthinspace10^{5}\!\times3$} & \textcolor{blue}{\scriptsize{}$220\negthinspace\times\negthinspace500$} & \textcolor{blue}{\scriptsize{}$15\negthinspace\times\negthinspace380\!\times80$}\tabularnewline
\emph{\footnotesize{}C} & \textcolor{blue}{\scriptsize{}$12\negthinspace\times\negthinspace6.9\negthinspace\cdot\negthinspace10^{5}\!\times3$} & \textcolor{blue}{\scriptsize{}$1\negthinspace\times\negthinspace9.7\negthinspace\cdot\negthinspace10^{5}\!\times3$} & \textcolor{blue}{\scriptsize{}$415\negthinspace\times\negthinspace1000$} & \textcolor{blue}{\scriptsize{}$35\negthinspace\times\negthinspace520\!\times110$} & \textcolor{blue}{\scriptsize{}$12\negthinspace\times\negthinspace5.9\negthinspace\cdot\negthinspace10^{5}\!\times4$} & \textcolor{blue}{\scriptsize{}$12\negthinspace\times\negthinspace5.6\negthinspace\cdot\negthinspace10^{5}\!\times4$} & \textcolor{blue}{\scriptsize{}$425\negthinspace\times\negthinspace1000$} & \textcolor{blue}{\scriptsize{}$36\negthinspace\times\negthinspace530\!\times110$}\tabularnewline
\emph{\footnotesize{}D} & \textcolor{blue}{\scriptsize{}$16\negthinspace\times\negthinspace1.8\negthinspace\cdot\negthinspace10^{6}\!\times4$} & \textcolor{blue}{\scriptsize{}$1\negthinspace\times\negthinspace1.8\negthinspace\cdot\negthinspace10^{6}\!\times4$} & \textcolor{blue}{\scriptsize{}$480\negthinspace\times\negthinspace2000$} & \textcolor{blue}{\scriptsize{}$55\negthinspace\times\negthinspace880\!\times180$} & \textcolor{blue}{\scriptsize{}$16\negthinspace\times\negthinspace1.5\negthinspace\cdot\negthinspace10^{6}\!\times5$} & \textcolor{blue}{\scriptsize{}$16\negthinspace\times\negthinspace1.5\negthinspace\cdot\negthinspace10^{6}\!\times5$} & \textcolor{blue}{\scriptsize{}$480\negthinspace\times\negthinspace2000$} & \textcolor{blue}{\scriptsize{}$55\negthinspace\times\negthinspace890\!\times180$}\tabularnewline
\hline 
\end{tabular}
\par\end{centering}{\footnotesize \par}

\caption{\label{tab:Configuration-parameters-1} Configuration parameters for
the BS HW model and for the Heston model, Static and Dynamic for the
GMWB-YD product with $\left(T_{1},T_{2}\right)=\left(10,25\right)$}
\end{table}

\subsection{\label{5.1}Static Withdrawal for GMWB-CF}

In the Static Withdrawal case we suppose the PH to withdrawal exactly
at the guaranteed rate.

The Static Tests 1 and 2 are inspired by \cite{CF}: in their article,
Chen and Forsyth price a GMWB contract according to an optimal withdrawal
framework, under the Black Scholes model. First we priced their product
for different maturities and withdrawal rates, assuming Static withdrawals
in Black and Scholes model to get a reference price in this model;
we got the $\alpha$ value using both a standard Monte Carlo method
and a standard PDE method. As we easily got the correct values for
the simple Black-Scholes model, then we add stochastic interest rate
and stochastic volatility. Model parameters are available in Table
\ref{tab:BS-CF-para}, and the values of $\alpha_{g}$ that we got
in the Black-Scholes case are given in Table \ref{tab:BS-CF-alpha}.

\subsubsection{Test 1: Static GMWB-CF in the Black-Scholes Hull-White Model}

In this test we want to price a GMWB-CF product according to BS HW
model. We use the same corresponding parameters as in the Black Scholes
model. Model parameters are shown in Table \ref{tab:mp1}. Results
are available in Table \ref{tab:Test1}. 

All the four methods behaved well and in the configuration D they
gave results consistent with the benchmark. HPDE proved to be the
best: all of its configurations gave results consistent with the benchmark.
Then APDE and SMC, and HMC gave good results too. SMC performed a
little better than HMC: the first method simulates the underlying
value and the interest rate exactly and so it is enough to simulate
the values at each event time. HMC matches the first three moments
of the BS HW $r$ process, but it doesn't reproduce exactly its law:
therefore it is right to increase the number of steps per year. So,
for a given run time, we can simulate less scenarios in HMC than SMC:
effectively, the confidence interval of HMC is larger than SMC one.
Moreover, SMC over performed the benchmark when using configuration
D. The two PDE methods returned stable results, and they often converged
in a monotone way.

With regard to the numerical results, we observe that the $\alpha_{g}$
values decrease with increasing maturity, just as in the Black-Scholes
case, and increase a little, with increasing withdrawal frequency.

\subsubsection{Test 2: Static GMWB-CF in the Heston Model}

In this test we want to price a GMWB-CF product according to the Heston
model. Model parameters are shown in the Table \ref{tab:mp2}. Results
are shown in Table \ref{tab:Test2}.

In this Test, MC methods had more difficulties; all the values of
PDE methods were close to the benchmark, while some values from MC
methods were less accurate, but compatibles with the benchmark (the
value of BM is inside the MC confidence interval). If we compare the
two MC approaches, in this case they are equivalent: HMC proved to
be faster than SMC when using few time steps (we could exploit $+11\%$
simulations in configuration A), while SMC proved to be slightly faster
in high time steps simulations, because of more time needed to build
the volatility tree ($-8\%$ simulations in configuration D). HPDE
shows to be very stable (case $T_{2}=10$, $WF=2$, $\alpha_{g}$
didn't change through configurations B-D), APDE behaved well to (often
monotone convergence). 

With regard to the numerical results, we observe that the $\alpha_{g}$
values decrease with increasing maturity, just as in the Black-Scholes
case, and increase a little, with increasing withdrawal frequency.

\subsubsection{Test 3: Hedging for Static GMWB-CF}

To reduce financial risks, insurance companies have to hedge the sold
VA: to accomplish this target they must calculate the Greeks of products.

In this test we want to show how the different methods can be used
to calculate the main Greeks. This can be done through finite differences
for small shocks on the variables. In general, the PDE methods are
ahead w.r.t. MC methods: the price is computed through finite differences
and so the price under shock is already computed. For MC methods this
is quite harder because the pricing has to be repeated changing the
inputs.

To start, we calculate the underlying greek Delta, for the products
of Test 1 and Test 2. As in this case we don't want to compute the
fair fee $\alpha_{g}$, we fix it arbitrarily: see Table \ref{tab:alpha-1}
and Table \ref{tab:alpha-2}. The values chosen are such as to cover
the costs of the insurer, and may be plausible on a real case. Results
are available in Table \ref{tab:Test3-1} (all values in Table must
be multiplied by $10^{-4}$).

In this Test, we got very accurate results with all method. Anyway,
HPDE and APDE proved to be the best: they both gave stable and accurate
results; in this Test, the two PDE methods were equivalent. We remark
that despite fair fee changes a lot when changing the maturity of
the policy, the value of Delta changes much less. Delta calculation
proved to be slightly harder in the Heston model case than in the
BS HW model case: see confidence intervals.

\clearpage

\begin{table}
\begin{centering}
\begin{tabular}{|lr|r|rr|}
{\small{}Expiry Time $T$} & {\small{}5, 10, 20 Years} &  & {\small{}GMW $G$} & {\small{}$100.0/(T\cdot WF)$}\tabularnewline
{\small{}Withdrawal Frequency $WF$} & {\small{}1 or 2 per Year} &  & {\small{}Initial Premium} & {\small{}$100.0$}\tabularnewline
{\small{}Initial account value $A_{0}$} & {\small{}$100.0$} &  & {\small{}$S_{0}$} & {\small{}$100.0$}\tabularnewline
{\small{}Initial base b. value $B_{0}$} & {\small{}$100.0$} &  & {\small{}$r$} & {\small{}$0.05$}\tabularnewline
{\small{}Withdrawal penalty $\kappa$} & {\small{}$0.10$} &  & {\small{}$\sigma$} & {\small{}$0.20$}\tabularnewline
Management fees $\alpha_{m}$ & $0$ &  &  & \tabularnewline
\end{tabular}
\par\end{centering}

\caption{\label{tab:BS-CF-para}Parameters used by Chen and Forsyth in \cite{CF}.}

\vspace{1cm}

\begin{centering}
\begin{tabular}{c|cc|cc|}
\multirow{2}{*}{{\small{}$T_{2}$}} & \multicolumn{2}{c|}{\emph{\small{}$WF=1$}} & \multicolumn{2}{c|}{\emph{\small{}$WF=2$}}\tabularnewline
 & {\small{}PDE} & {\small{}MC} & {\small{}PDE} & {\small{}MC}\tabularnewline
\hline 
{\small{}$5$} & {\footnotesize{}$235.24$} & {\footnotesize{}$235.11\pm0.42$} & {\footnotesize{}$243.96$} & {\footnotesize{}$243.80\pm0.42$}\tabularnewline
{\small{}$10$} & {\footnotesize{}$92.41$} & {\footnotesize{}$92.28\pm0.30$} & {\footnotesize{}$94.62$} & {\footnotesize{}$94.84\pm0.30$}\tabularnewline
{\small{}$20$} & {\footnotesize{}$27.64$} & {\footnotesize{}$27.79\pm0.24$} & {\footnotesize{}$28.09$} & {\footnotesize{}$28.39\pm0.24$}\tabularnewline
\end{tabular}
\par\end{centering}

\caption{\label{tab:BS-CF-alpha}Fair bp values of $\alpha_{g}$ in Black Scholes
model, for Static GMWB-CF with the same parameters as in \cite{CF}.}
\end{table}

\begin{table}
\begin{centering}
{\small{}}%
\begin{tabular}{c|ccccc|c}
\hline 
{\small{}$S_{0}$} & {\small{}$r$} & {\small{}$curve$} & {\small{}$k$} & {\small{}$\omega$} & {\small{}$\rho$} & {\small{}$\sigma$}\tabularnewline
{\small{}$100$} & {\small{}$0.05$} & {\small{}$flat$} & {\small{}$1.0$} & {\small{}$0.2$} & {\small{}$-0.5$} & {\small{}$0.20$}\tabularnewline
\hline 
\end{tabular}
\par\end{centering}{\small \par}

\caption{\label{tab:mp1}The model parameters about Test 1}

\vspace{1cm}

\begin{centering}
%
\par\end{centering}

\caption{\label{tab:Test1}Test 1. In the first Table, the fair fee $\alpha_{g}$
in bps for the Black-Scholes Hull-White model, with annual or six-monthly
withdrawal. In the second Table the run times for the case $WF=1$,
$T_{2}=10$. Finally, the plot of relative error (w.r.t. BM value)
for the four methods in the same case of run-times Table. The parameters
used for this test are available in Table \ref{tab:BS-CF-para} and
in Table \ref{tab:mp1}.}
\end{table}

\begin{table}
\begin{centering}
{\small{}}%
\begin{tabular}{c|ccccc|c}
\hline 
{\small{}$S_{0}$} & {\small{}$v_{0}$} & {\small{}$\theta$} & {\small{}$k$} & {\small{}$\omega$} & {\small{}$\rho$} & {\small{}$r$}\tabularnewline
{\small{}$100$} & {\small{}$0.20^{2}$} & {\small{}$0.20^{2}$} & {\small{}$1.0$} & {\small{}$0.2$} & {\small{}$-0.5$} & {\small{}$0.05$}\tabularnewline
\hline 
\end{tabular}
\par\end{centering}{\small \par}

\caption{\label{tab:mp2}The model parameters about Test 2.}

\vspace{1cm}

\begin{centering}
%
\par\end{centering}

\caption{\label{tab:Test2}Test 2. In the first Table, the fair fee $\alpha_{g}$
in bps for the Heston model, with annual or six-monthly withdrawal.
In the second Table the run times for the case $WF=1$, $T_{2}=10$.
Finally, the plot of relative error (w.r.t. BM value) for the four
methods in the same case of run-times Table. The parameters used for
this test are available in Table \ref{tab:BS-CF-para} and in Table
\ref{tab:mp2}.}
\end{table}

\begin{table}
\begin{centering}
{\small{}}%
\begin{tabular}{c|cc}
\hline 
{\small{}$T_{2}$} & \emph{\small{}$WF=1$} & \emph{\small{}$WF=2$}\tabularnewline
\hline 
{\small{}$5$} & {\scriptsize{}$200$} & {\scriptsize{}$200$}\tabularnewline
{\small{}$10$} & {\scriptsize{}$100$} & {\scriptsize{}$100$}\tabularnewline
{\small{}$20$} & {\scriptsize{}$50$} & {\scriptsize{}$50$}\tabularnewline
\hline 
\end{tabular}
\par\end{centering}{\small \par}

\caption{\label{tab:alpha-1}The $\alpha_{g}$ values used for Delta calculation
in the Static BS HW case (bps).}

\vspace{0.3cm}

\begin{centering}
{\small{}}%
\begin{tabular}{|cc|ccccc||ccccc|}
\hline 
\multirow{2}{*}{{\small{}$T_{2}$}} & \multirow{2}{*}{} & \multicolumn{5}{c||}{\emph{\small{}$WF=1$}} & \multicolumn{5}{c|}{\emph{\small{}$WF=2$}}\tabularnewline
\cline{3-12} 
 &  & {\footnotesize{}HMC} & {\footnotesize{}SMC} & {\footnotesize{}HPDE} & {\footnotesize{}APDE} & {\footnotesize{}BM} & {\footnotesize{}HMC} & {\footnotesize{}SMC} & {\footnotesize{}HPDE} & {\footnotesize{}APDE} & {\footnotesize{}BM}\tabularnewline
\hline 
\multirow{4}{*}{{\small{}$5$}} & \emph{\small{}A} & \textcolor{blue}{\scriptsize{}$6212\pm\phantom{0}4$} & \textcolor{blue}{\scriptsize{}$6214\pm\phantom{0}3$} & \textcolor{blue}{\scriptsize{}$6213$} & \textcolor{blue}{\scriptsize{}$6212$} & \multirow{2}{*}{\textcolor{blue}{\scriptsize{}$6213$}} & \textcolor{blue}{\scriptsize{}$6178\pm\phantom{0}4$} & \textcolor{blue}{\scriptsize{}$6180\pm\phantom{0}4$} & \textcolor{blue}{\scriptsize{}$6181$} & \textcolor{blue}{\scriptsize{}$6180$} & \multirow{2}{*}{\textcolor{blue}{\scriptsize{}$6180$}}\tabularnewline
 & \emph{\small{}B} & \textcolor{blue}{\scriptsize{}$6213\pm\phantom{0}3$} & \textcolor{blue}{\scriptsize{}$6213\pm\phantom{0}1$} & \textcolor{blue}{\scriptsize{}$6213$} & \textcolor{blue}{\scriptsize{}$6213$} &  & \textcolor{blue}{\scriptsize{}$6180\pm\phantom{0}3$} & \textcolor{blue}{\scriptsize{}$6180\pm\phantom{0}2$} & \textcolor{blue}{\scriptsize{}$6180$} & \textcolor{blue}{\scriptsize{}$6180$} & \tabularnewline
 & \emph{\small{}C} & \textcolor{blue}{\scriptsize{}$6211\pm\phantom{0}1$} & \textcolor{blue}{\scriptsize{}$6213\pm\phantom{0}1$} & \textcolor{blue}{\scriptsize{}$6213$} & \textcolor{blue}{\scriptsize{}$6213$} & \multirow{2}{*}{\textcolor{blue}{\scriptsize{}$\pm1$}} & \textcolor{blue}{\scriptsize{}$6179\pm\phantom{0}1$} & \textcolor{blue}{\scriptsize{}$6180\pm\phantom{0}1$} & \textcolor{blue}{\scriptsize{}$6180$} & \textcolor{blue}{\scriptsize{}$6180$} & \multirow{2}{*}{\textcolor{blue}{\scriptsize{}$\pm1$}}\tabularnewline
\hline 
\end{tabular}
\par\end{centering}{\small \par}

\caption{\label{tab:Test3-1}Test 3. Delta calculation for the Static BS HW
case. All results must be multiplied by $10^{-4}$. The parameters
used for this test are available in Table \ref{tab:BS-CF-para}, \ref{tab:mp2}
and in Table \ref{tab:alpha-1} .}
\end{table}

\begin{table}
\begin{centering}
{\small{}}%
\begin{tabular}{c|cc}
\hline 
{\small{}$T_{2}$} & \emph{\small{}$WF=1$} & \emph{\small{}$WF=2$}\tabularnewline
\hline 
{\small{}$5$} & {\scriptsize{}$250$} & {\scriptsize{}$250$}\tabularnewline
{\small{}$10$} & {\scriptsize{}$100$} & {\scriptsize{}$100$}\tabularnewline
{\small{}$20$} & {\scriptsize{}$50$} & {\scriptsize{}$50$}\tabularnewline
\hline 
\end{tabular}
\par\end{centering}{\small \par}

\caption{\label{tab:alpha-2}The $\alpha_{g}$ values used for Delta calculation
in the Static Heston case (bps).}

\vspace{0.3cm}

\begin{centering}
{\small{}}%
\begin{tabular}{|cc|ccccc||ccccc|}
\hline 
\multirow{2}{*}{{\small{}$T_{2}$}} & \multirow{2}{*}{} & \multicolumn{5}{c||}{\emph{\small{}$WF=1$}} & \multicolumn{5}{c|}{\emph{\small{}$WF=2$}}\tabularnewline
\cline{3-12} 
 &  & {\footnotesize{}HMC} & {\footnotesize{}SMC} & {\footnotesize{}HPDE} & {\footnotesize{}APDE} & {\footnotesize{}BM} & {\footnotesize{}HMC} & {\footnotesize{}SMC} & {\footnotesize{}HPDE} & {\footnotesize{}APDE} & {\footnotesize{}BM}\tabularnewline
\hline 
\multirow{4}{*}{{\small{}$5$}} & \emph{\small{}A} & \textcolor{blue}{\scriptsize{}$6132\pm\phantom{0}4$} & \textcolor{blue}{\scriptsize{}$6141\pm\phantom{0}5$} & \textcolor{blue}{\scriptsize{}$6132$} & \textcolor{blue}{\scriptsize{}$6131$} & \multirow{2}{*}{\textcolor{blue}{\scriptsize{}$6131$}} & \textcolor{blue}{\scriptsize{}$6101\pm\phantom{0}5$} & \textcolor{blue}{\scriptsize{}$6107\pm\phantom{0}5$} & \textcolor{blue}{\scriptsize{}$6099$} & \textcolor{blue}{\scriptsize{}$6098$} & \multirow{2}{*}{\textcolor{blue}{\scriptsize{}$6098$}}\tabularnewline
 & \emph{\small{}B} & \textcolor{blue}{\scriptsize{}$6134\pm\phantom{0}3$} & \textcolor{blue}{\scriptsize{}$6136\pm\phantom{0}3$} & \textcolor{blue}{\scriptsize{}$6131$} & \textcolor{blue}{\scriptsize{}$6131$} &  & \textcolor{blue}{\scriptsize{}$6101\pm\phantom{0}3$} & \textcolor{blue}{\scriptsize{}$6104\pm\phantom{0}3$} & \textcolor{blue}{\scriptsize{}$6098$} & \textcolor{blue}{\scriptsize{}$6098$} & \tabularnewline
 & \emph{\small{}C} & \textcolor{blue}{\scriptsize{}$6131\pm\phantom{0}2$} & \textcolor{blue}{\scriptsize{}$6131\pm\phantom{0}2$} & \textcolor{blue}{\scriptsize{}$6131$} & \textcolor{blue}{\scriptsize{}$6131$} & \multirow{2}{*}{\textcolor{blue}{\scriptsize{}$\pm1$}} & \textcolor{blue}{\scriptsize{}$6099\pm\phantom{0}2$} & \textcolor{blue}{\scriptsize{}$6097\pm\phantom{0}2$} & \textcolor{blue}{\scriptsize{}$6098$} & \textcolor{blue}{\scriptsize{}$6098$} & \multirow{2}{*}{\textcolor{blue}{\scriptsize{}$\pm1$}}\tabularnewline
\hline 
\end{tabular}
\par\end{centering}{\small \par}

\caption{\label{tab:Test3-2}Test 3. Delta calculation for the Static Heston
case. All results must be multiplied by $10^{-4}$. The parameters
used for this test are available in Table \ref{tab:BS-CF-para}, \ref{tab:mp2}
and in Table \ref{tab:alpha-2} .}
\end{table}

\clearpage

\subsection{\label{5.2}Dynamic Withdrawal for GMWB-CF}

In the Dynamic withdrawal case we suppose the PH to chose at each
event time how much withdraw, in order to maximize his (her) gain
(optimal withdrawal).

The Static Tests 4 and 5 are inspired by \cite{CF}: in their article,
Chen and Forsyth price a GMWB contract in a optimal withdrawal framework,
under the Black Scholes model. First we priced their product for different
maturities and withdrawal rates, assuming optimal withdrawals in Black
and Scholes model to get a reference price in this model; we got the
$\alpha$ value using both a Regression by Lines Monte Carlo method
and a standard PDE method. As we got the good values for the simple
Black-Scholes model, then we add stochastic interest rate and stochastic
volatility. Model parameters are available in Table \ref{tab:BS-CF-para},
and the values of $\alpha_{g}$ that we got are given in Table \ref{tab:BS-CF-alpha-2}.

We remark that we used the Full Regression algorithm for the calculation
of the MC prices (case A, B, C, D for SMC and HMC): this method is
quite fast, however the results quality is low.

Conversely, we used the Regression by Lines algorithm to calculate
the benchmarks (BM): this algorithm is much more time demanding than
the Full Regression, but its results are higher, proving that the
regression performs better and the PH, using this approach, can have
a better payoff. Moreover, this method performed very well in the
Black-Scholes model, and we used it to fill Table \ref{tab:BS-CF-alpha-2}.
We tried to use Regression by Lines algorithm also for cases A, B,
C, D but we didn't get good results, because of the short run time
available (max 30 mins).

For benchmarks calculation, we used 4 degree polynomials with $10^{6}$
scenarios (doubled by the antithetic variables technique), excluding
the case $T_{2}=20$, $WF=2$ where we used half scenarios: the time
needed to perform these calculations (two secant steps around the
value of case D of HPDE) varies from 30 minutes (case $T_{2}=5$ ,
$WF=1$) to 38 hours (case $T_{2}=20$, $WF=2$). 

We would remark that, using PDE method for the Black Scholes model,
we obtained the same values as in \cite{CF} (only two values are
available in Chen and Forsyth's paper), but MC method (Regression
by Lines) had a few problems (lower values): the least squares regression
doesn't work very well and this problem is stiff for MC methods (see
Table \ref{tab:BS-CF-alpha-2}). We can therefore imagine that the
MC methods will have difficulties also in the following tests, in
which a dimension is added. 

\begin{table}
\begin{centering}
\begin{tabular}{c|ccc|ccc|}
\multirow{2}{*}{{\small{}$T_{2}$}} & \multicolumn{3}{c|}{\emph{\small{}$WF=1$}} & \multicolumn{3}{c|}{\emph{\small{}$WF=2$}}\tabularnewline
 & {\small{}PDE} & {\small{}MC} & Ch.Fo. & {\small{}PDE} & {\small{}MC} & Ch.Fo.\tabularnewline
\hline 
{\small{}$5$} & {\footnotesize{}$248.33$} & {\footnotesize{}$247.75\pm1.39$} & $n.c.$ & {\footnotesize{}$258.20$} & {\footnotesize{}$257.32\pm1.42$} & $n.c.$\tabularnewline
{\small{}$10$} & {\footnotesize{}$129.18$} & {\footnotesize{}$128.58\pm1.08$} & {\footnotesize{}$129.10$} & {\footnotesize{}$133.60$} & {\footnotesize{}$133.09\pm1.11$} & {\footnotesize{}$133.52$}\tabularnewline
{\small{}$20$} & {\footnotesize{}$66.42$} & {\footnotesize{}$66.20\pm0.89$} & $n.c.$ & {\footnotesize{}$68.59$} & {\footnotesize{}$68.52\pm1.29$} & $n.c.$\tabularnewline
\end{tabular}
\par\end{centering}

\caption{\label{tab:BS-CF-alpha-2}Fair bp values of $\alpha_{g}$ in Black
Scholes model, for Dynamic GMWB-CF with the same parameters as in
\cite{CF}. The values that aren't available in \cite{CF} (not computed)
are denoted by \emph{``n.c.''} .}
\end{table}

\subsubsection{Test 4: Dynamic GMWB-CF in the Black-Scholes Hull-White Model}

Test 4 is the Dynamic case of Test 1. Model parameters are shown in
Table \ref{tab:mp1}. Results are available in Table \ref{tab:Test4}.
In this test PDE methods proved to be much more efficient than MC
ones. In fact MC methods use a least-squares regression approach to
find the optimal withdrawal: this method needs a lot of scenarios
to approximate through the regression the value of the policy for
a given set of variable, and this is time demanding. Then, working
at fixed time, we could perform fewer scenarios than the Static case.
PDE methods felt another problem: the increase of problem dimension
forced us to reduce the number of time steps wrt Static case. Using
MC methods, we always got lower values with regard to PDE methods,
and moreover MC values increased by several bps when moving from configuration
A to D.

The two MC methods proved to be equivalent: the differences in scenarios
generation run-time are negligible because most of the time is spent
in finding the best withdrawal. The HPDE method gave good and stable
results, while APDE had more troubles, with results floating around
the good values. Then, the HPDE method proved to be the best one according
to the results of this test.

The case $\left(T_{2},WF\right)=\left(20,2\right)$ proved to be very
insidious: the long maturity and the large number of withdrawal dates
(40 event times) made the problem hard also for PDE methods. In this
case MC methods in configuration A also gave lower values than Static
approach ($18.64$ vs $25.20$): due to the few scenarios considered,
the least squares regression failed to increase PH's gain.

\subsubsection{Test 5: Dynamic GMWB-CF in the Heston Model}

Test 5 is the Dynamic case of Test 2. Model parameters are shown in
Table \ref{tab:mp2}. Results are available in Table \ref{tab:Test5}.

In this test things are similar to Test 4, but the optimization problem
seemed to be easier than in Test 4: MC methods converged better, especially
when using high level configurations. PDE methods behaved good as
usual, and in this case they proved to be almost equivalent: they
both gave good results except for the case $\left(T_{2},WF\right)=\left(20,2\right)$
where the initial results of APDE were too high. The two MC methods
proved to be equivalent. We note that, in the Heston model case, Dynamic
strategy increased the value of $\alpha_{g}$ less than in BS HW case:
probably, playing on interest rate lets the PH gain more than playing
on volatility.

The case $\left(T_{2},WF\right)=\left(20,2\right)$ is still the most
insidious, but this time we didn't get any value lower than the Static
value of $\alpha_{g}$.

\subsubsection{Test 6: Hedging for Dynamic GMWB-CF }

Test 6 is the Dynamic case of Test 3. Results are available in Table
\ref{tab:Test6}.

In this test we got good results with PDE methods: values of HPDE
are very regular despite the high dimension of the problem. Results
from APDE are good, but a bit worse than HPDE especially in BS HW
case (see for example case $\left(T_{2},WF\right)=\left(20,2\right)$).
Monte Carlo methods suffered the few scenarios performed and sometimes
the confidence interval is very large. In the case $\left(T_{2},WF\right)=\left(20,2\right)$
we also got some convergence problems in the BS HW model.

\subsubsection{Optimal Withdrawal Strategy Plots for Dynamic GMWB-CF }

\begin{figure}
\begin{centering}
\includegraphics[scale=0.14]{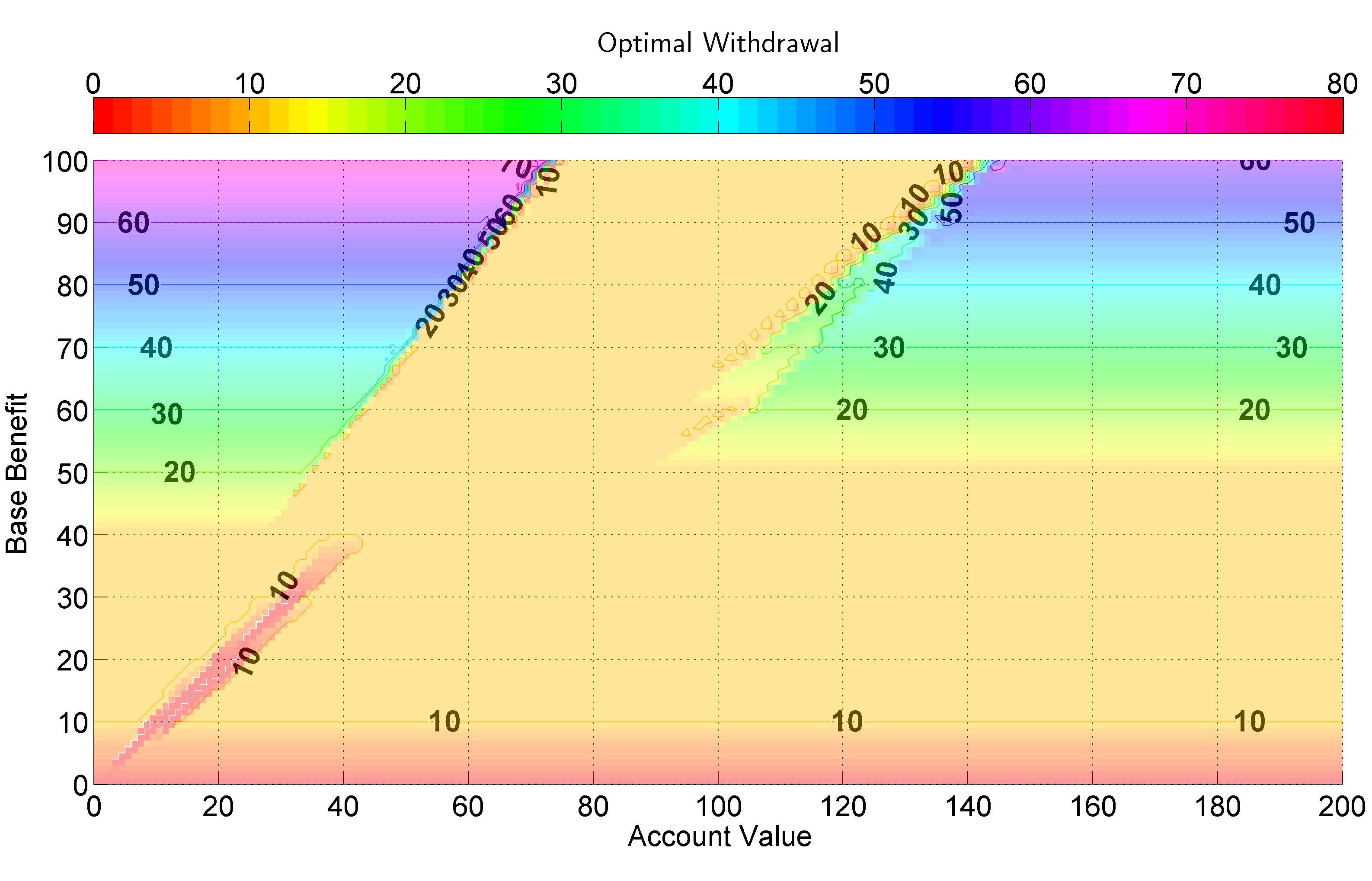}
\par\end{centering}

\vspace{0.1cm}

\begin{centering}
\includegraphics[scale=0.14]{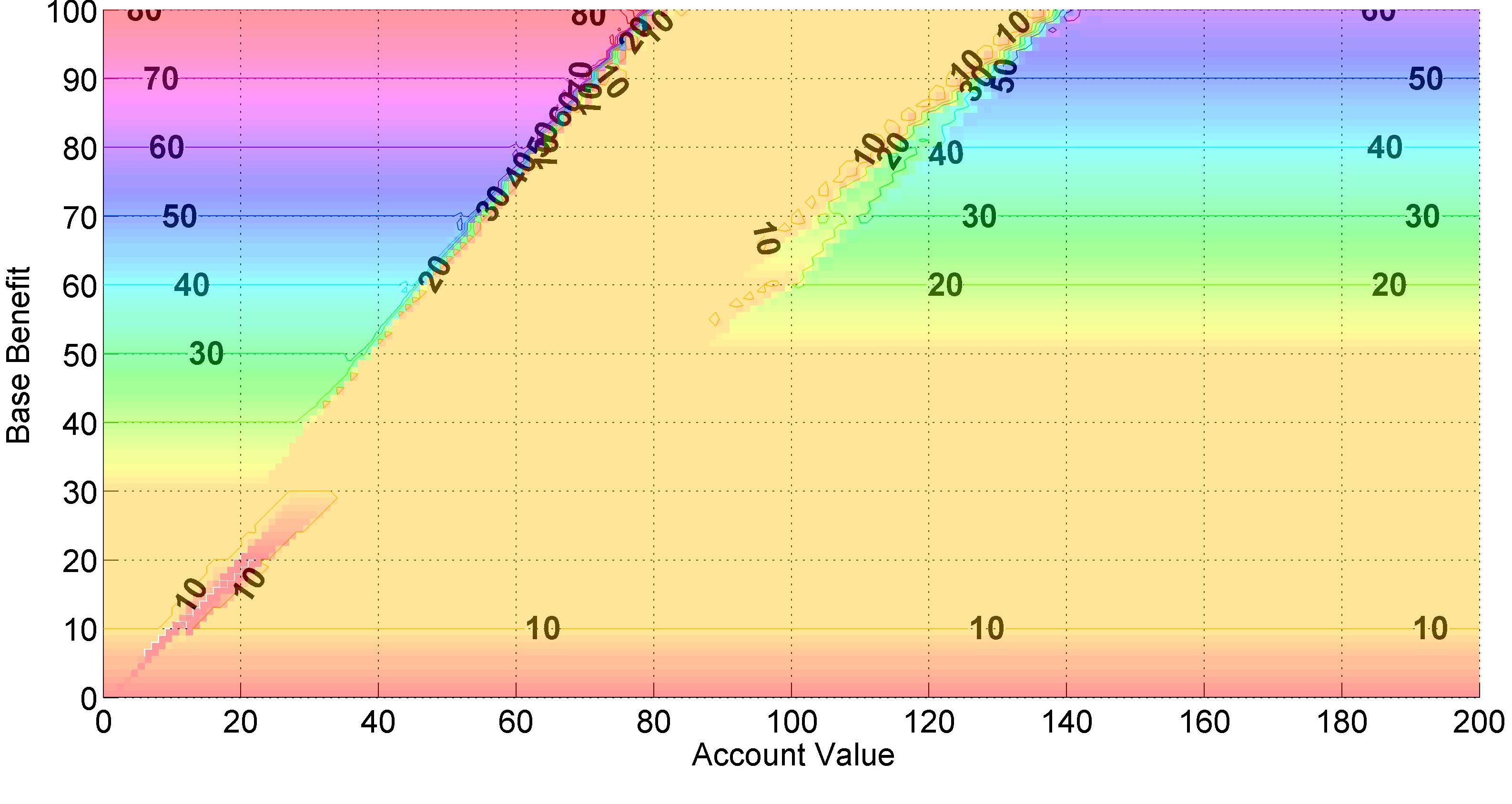}
\par\end{centering}

\vspace{0.1cm}

\begin{centering}
\includegraphics[scale=0.14]{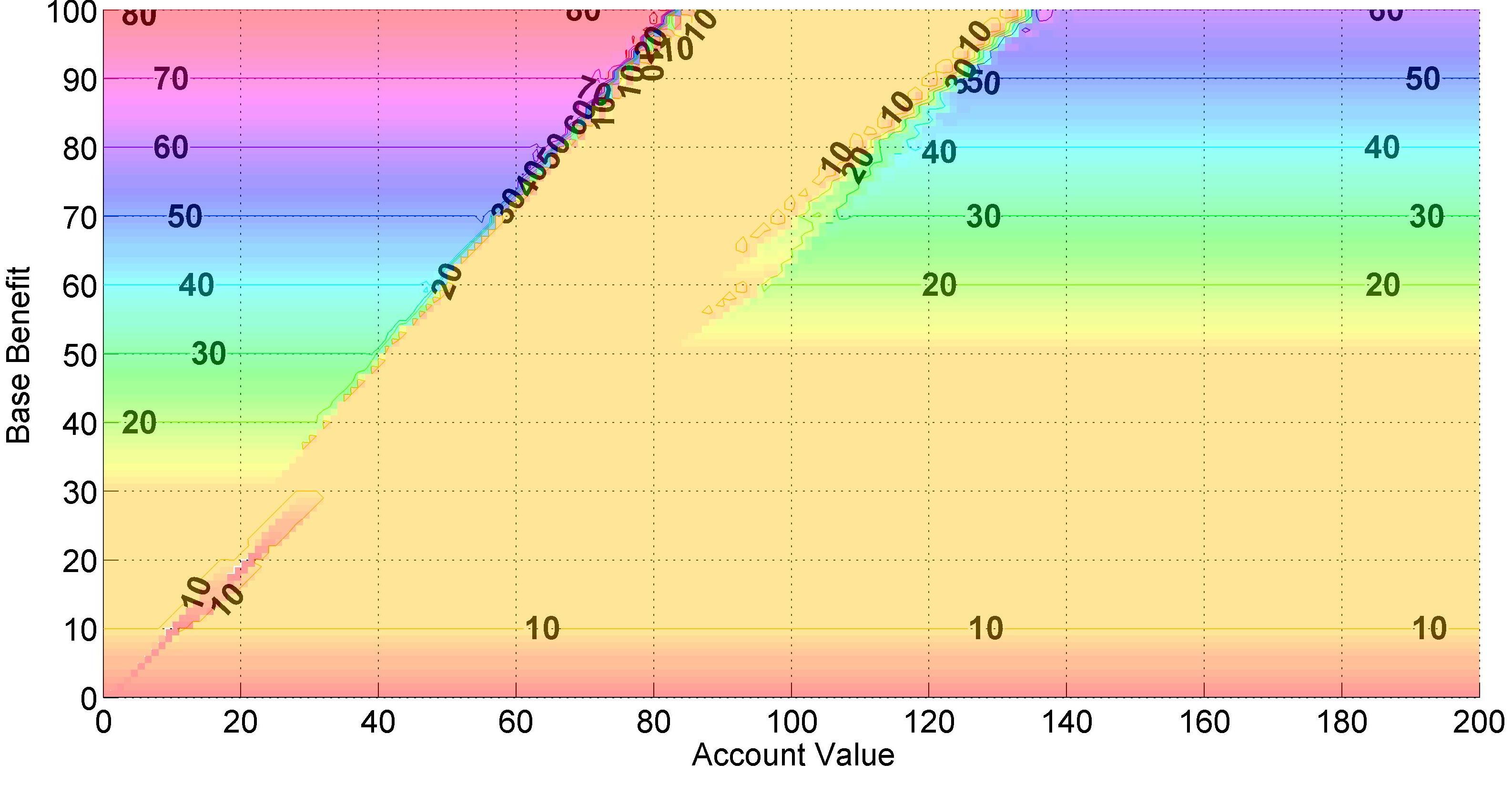}
\par\end{centering}

\caption{\label{fig:Plots-ow_HW}Plots of the optimal withdrawals at time $t=1$
for the BS HW model according to different values of $r_{1}$: from
the top to the bottom $r_{1}=0.03$, $r_{1}=0.05$ and $r_{1}=0.07$.
The parameters used to obtain these plots are the same as for Delta
calculation for case $T_{2}=10$, $WDF=1$: see Tables \ref{tab:BS-CF-para},
\ref{tab:mp1} and \ref{tab:alpha-3} .}

\end{figure}

\begin{figure}
\begin{centering}
\includegraphics[scale=0.14]{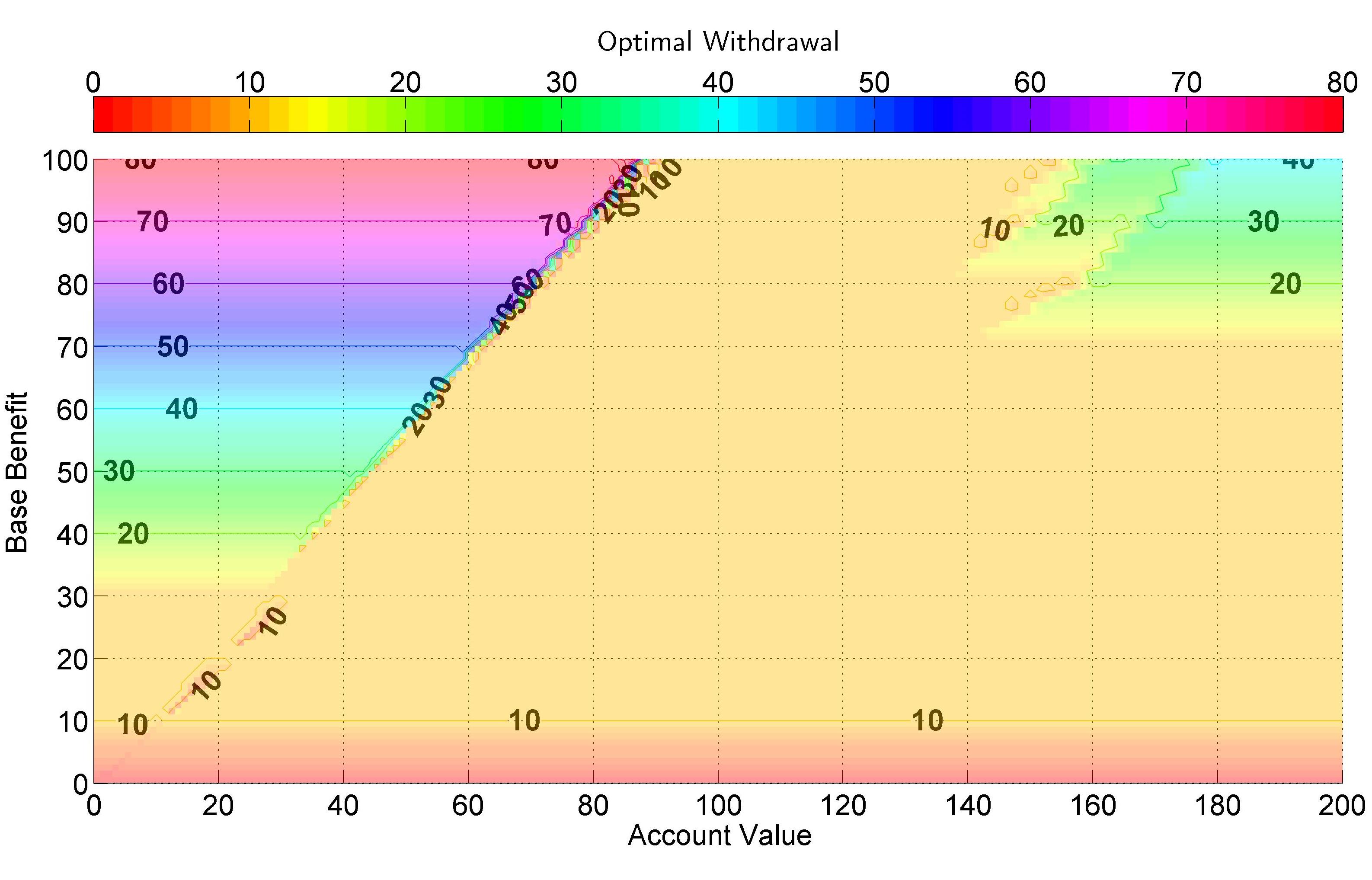}
\par\end{centering}

\vspace{0.1cm}

\begin{centering}
\includegraphics[scale=0.14]{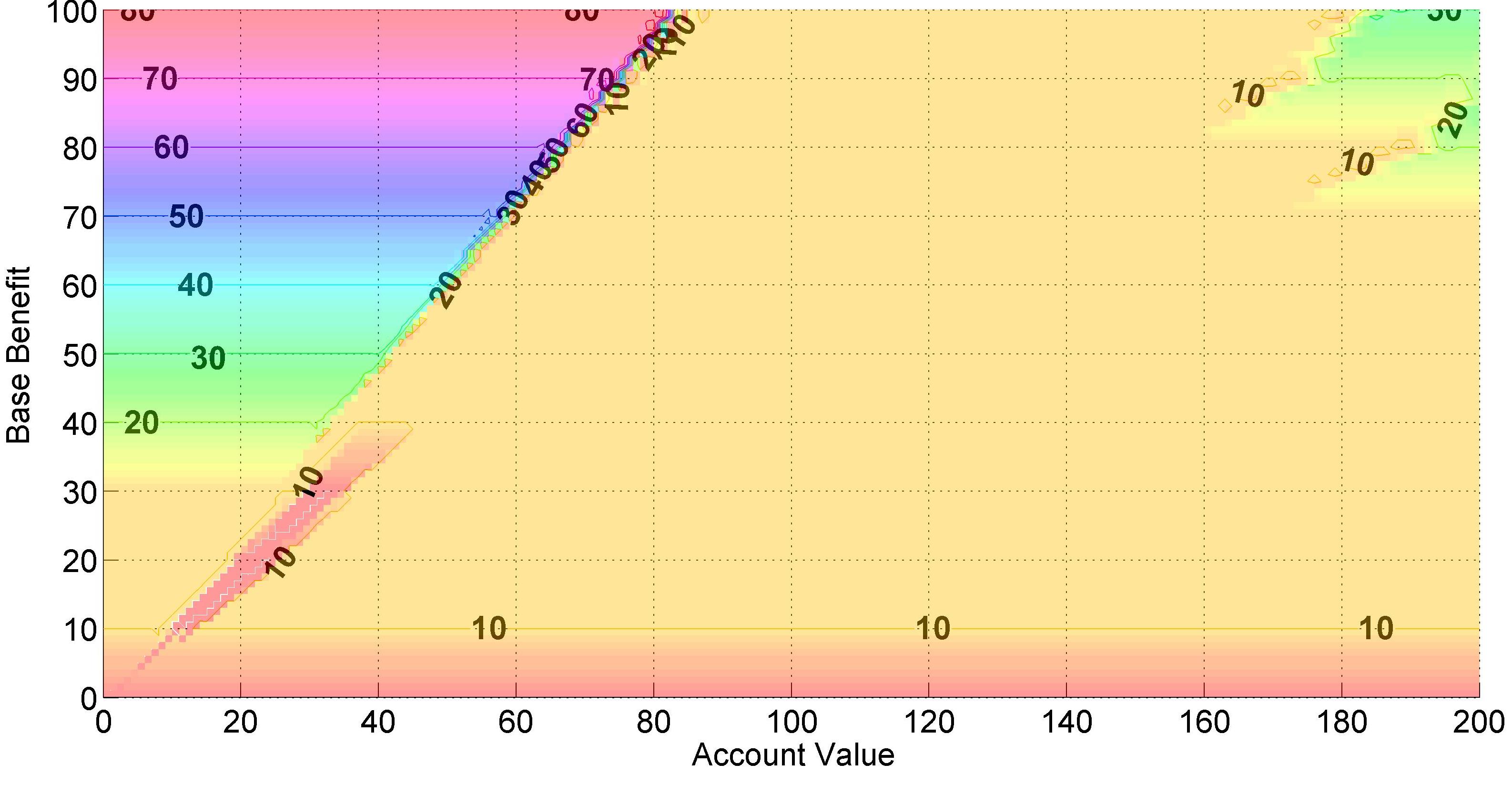}
\par\end{centering}

\vspace{0.1cm}

\begin{centering}
\includegraphics[scale=0.14]{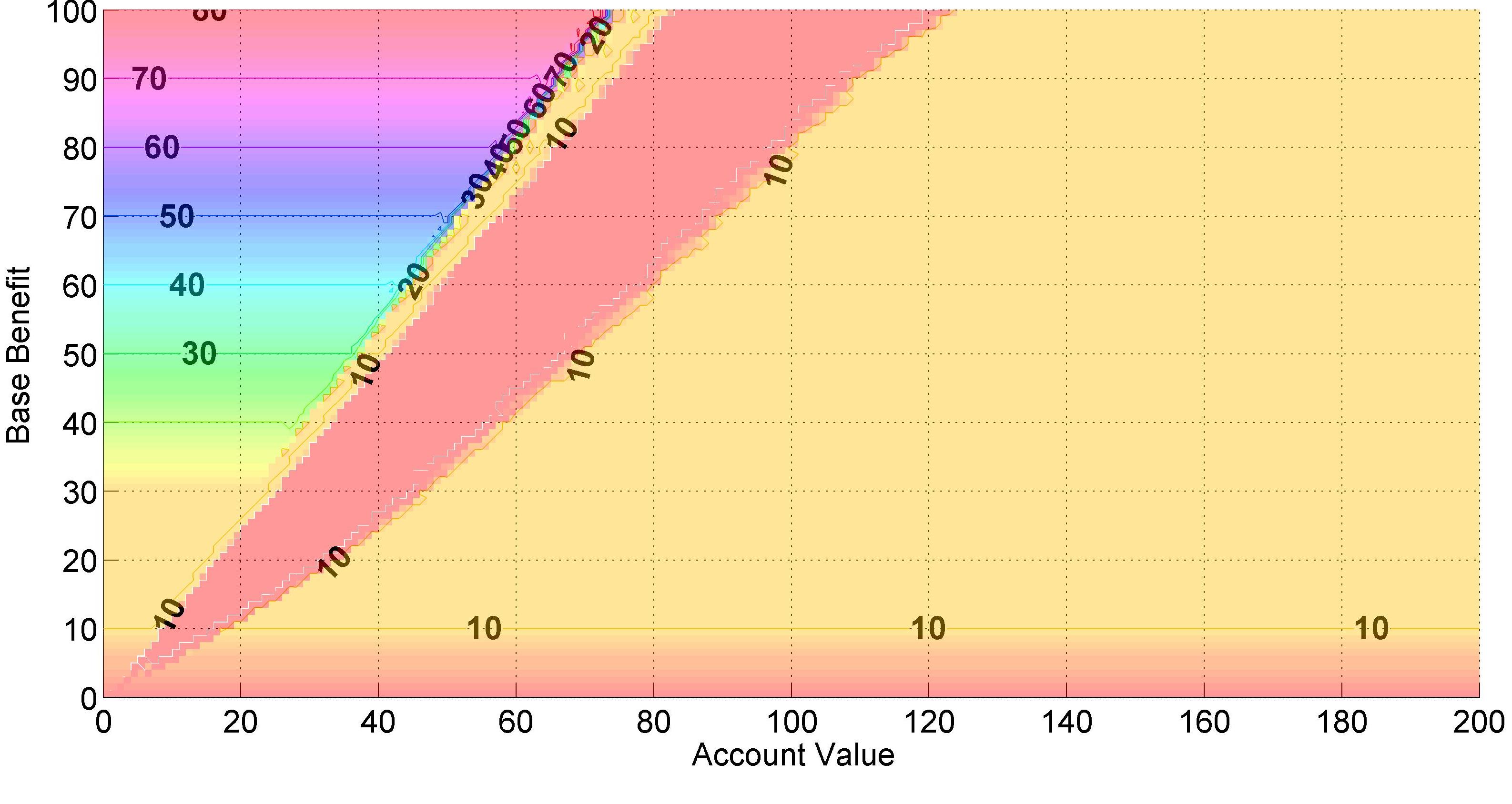}
\par\end{centering}

\caption{\label{fig:Plots-ow_He}Plots of the optimal withdrawals at time $t=1$
for the Heston model according to different values of the volatility
$v_{1}$: from the top to the bottom $v_{1}=0$, $v_{1}=0.04$ and
$v_{1}=0.16$. The parameters used to obtain these plots are the same
as for Delta calculation for case $T_{2}=10$, $WDF=1$: see Tables
\ref{tab:BS-CF-para}, \ref{tab:mp2} and \ref{tab:alpha-4}.}
\end{figure}

In Figure \ref{fig:Plots-ow_HW} and \ref{fig:Plots-ow_He} we calculated
the optimal withdrawal for the GMWB-CF product with $\left(T_{2},WF\right)=\left(10,1\right)$
for both the BS HW model and Heston model. We used HPDE methods to
obtain these plots: we chose three nodes of the tree around the initial
value at time $t=1$ and we used the best withdrawals to get these
plots.

We remark that these plots are very similar to those proposed in \cite{CF}:
we note the same structure around the bisector and the wide region
of regular withdrawal.

\clearpage

\begin{table}
\begin{centering}

\par\end{centering}

\caption{\label{tab:Test4}Test 4. In the first Table, the fair fee $\alpha_{g}$
in bps for the Black-Scholes Hull-White model, with annual or six-monthly
withdrawal. In the second Table the run times for the case $WF=1$
, $T_{2}=10$. Finally, the plot of relative error (w.r.t. BM value)
for the four methods in the same case of run-times Table. The parameters
used for this test are available in Table \ref{tab:BS-CF-para} and
in Table \ref{tab:mp1}.}
\end{table}

\begin{table}
\begin{centering}
%
\par\end{centering}

\caption{\label{tab:Test5}Test 5. In the first Table, the fair fee $\alpha_{g}$
in bps for the Heston model, with annual or six-monthly withdrawal.
In the second Table the run times for the case $WF=1$ , $T_{2}=10$.
Finally, the plot of relative error (w.r.t. BM value) for the four
methods in the same case of run-times Table. The parameters used for
this test are available in Table \ref{tab:BS-CF-para} and in Table
\ref{tab:mp2}.}
\end{table}

\begin{table}
\begin{centering}
{\small{}}%
\begin{tabular}{c|cc}
\hline 
{\small{}$T_{2}$} & \emph{\small{}$WF=1$} & \emph{\small{}$WF=2$}\tabularnewline
\hline 
{\small{}$5$} & {\scriptsize{}$350$} & {\scriptsize{}$350$}\tabularnewline
{\small{}$10$} & {\scriptsize{}$200$} & {\scriptsize{}$200$}\tabularnewline
{\small{}$20$} & {\scriptsize{}$150$} & {\scriptsize{}$150$}\tabularnewline
\hline 
\end{tabular}
\par\end{centering}{\small \par}

\caption{\label{tab:alpha-3}The $\alpha_{g}$ values used for Delta calculation
in the Dynamic BS HW case (bps).}

\vspace{0.3cm}

\begin{centering}
{\small{}}%
\begin{tabular}{|cc|ccccc||ccccc|}
\hline 
\multirow{2}{*}{{\small{}$T_{2}$}} & \multirow{2}{*}{} & \multicolumn{5}{c||}{\emph{\small{}$WF=1$}} & \multicolumn{5}{c|}{\emph{\small{}$WF=2$}}\tabularnewline
\cline{3-12} 
 &  & {\footnotesize{}HMC} & {\footnotesize{}SMC} & {\footnotesize{}HPDE} & {\footnotesize{}APDE} & {\footnotesize{}BM} & {\footnotesize{}HMC} & {\footnotesize{}SMC} & {\footnotesize{}HPDE} & {\footnotesize{}APDE} & {\footnotesize{}BM}\tabularnewline
\hline 
\multirow{4}{*}{{\small{}$5$}} & \emph{\small{}A} & \textcolor{blue}{\scriptsize{}$4986\pm311$} & \textcolor{blue}{\scriptsize{}$4790\pm310$} & \textcolor{blue}{\scriptsize{}$4474$} & \textcolor{blue}{\scriptsize{}$4465$} & \multirow{2}{*}{\textcolor{blue}{\scriptsize{}$4514$}} & \textcolor{blue}{\scriptsize{}$4753\pm537$} & \textcolor{blue}{\scriptsize{}$4117\pm429$} & \textcolor{blue}{\scriptsize{}$4191$} & \textcolor{blue}{\scriptsize{}$4187$} & \multirow{2}{*}{\textcolor{blue}{\scriptsize{}$4181$}}\tabularnewline
 & \emph{\small{}B} & \textcolor{blue}{\scriptsize{}$4455\pm171$} & \textcolor{blue}{\scriptsize{}$4443\pm179$} & \textcolor{blue}{\scriptsize{}$4477$} & \textcolor{blue}{\scriptsize{}$4469$} &  & \textcolor{blue}{\scriptsize{}$4220\pm182$} & \textcolor{blue}{\scriptsize{}$4362\pm222$} & \textcolor{blue}{\scriptsize{}$4196$} & \textcolor{blue}{\scriptsize{}$4191$} & \tabularnewline
 & \emph{\small{}C} & \textcolor{blue}{\scriptsize{}$4385\pm130$} & \textcolor{blue}{\scriptsize{}$4420\pm122$} & \textcolor{blue}{\scriptsize{}$4478$} & \textcolor{blue}{\scriptsize{}$4473$} & \multirow{2}{*}{\textcolor{blue}{\scriptsize{}$\pm62$}} & \textcolor{blue}{\scriptsize{}$4057\pm196$} & \textcolor{blue}{\scriptsize{}$3987\pm170$} & \textcolor{blue}{\scriptsize{}$4198$} & \textcolor{blue}{\scriptsize{}$4195$} & \multirow{2}{*}{\textcolor{blue}{\scriptsize{}$\pm68$}}\tabularnewline
\hline 
\end{tabular}
\par\end{centering}{\small \par}

\caption{\label{tab:Test6}Test 6. Delta calculation for the Dynamic BS HW
case. All results must be multiplied by $10^{-4}$. The parameters
used for this test are available in Table \ref{tab:BS-CF-para}, \ref{tab:alpha-3}
and in Table \ref{tab:mp2}.}
\end{table}

\begin{table}
\begin{centering}
{\small{}}%
\begin{tabular}{c|cc}
\hline 
{\small{}$T_{2}$} & \emph{\small{}$WF=1$} & \emph{\small{}$WF=2$}\tabularnewline
\hline 
{\small{}$5$} & {\scriptsize{}$300$} & {\scriptsize{}$300$}\tabularnewline
{\small{}$10$} & {\scriptsize{}$150$} & {\scriptsize{}$150$}\tabularnewline
{\small{}$20$} & {\scriptsize{}$100$} & {\scriptsize{}$100$}\tabularnewline
\hline 
\end{tabular}
\par\end{centering}{\small \par}

\caption{\label{tab:alpha-4}The $\alpha_{g}$ values used for Delta calculation
in the Dynamic Heston case (bps).}

\vspace{0.3cm}

\begin{centering}
{\small{}}%
\begin{tabular}{|cc|ccccc||ccccc|}
\hline 
\multirow{2}{*}{{\small{}$T_{2}$}} & \multirow{2}{*}{} & \multicolumn{5}{c||}{\emph{\small{}$WF=1$}} & \multicolumn{5}{c|}{\emph{\small{}$WF=2$}}\tabularnewline
\cline{3-12} 
 &  & {\footnotesize{}HMC} & {\footnotesize{}SMC} & {\footnotesize{}HPDE} & {\footnotesize{}APDE} & {\footnotesize{}BM} & {\footnotesize{}HMC} & {\footnotesize{}SMC} & {\footnotesize{}HPDE} & {\footnotesize{}APDE} & {\footnotesize{}BM}\tabularnewline
\hline 
\multirow{4}{*}{{\small{}$5$}} & \emph{\small{}A} & \textcolor{blue}{\scriptsize{}$5732\pm\phantom{0}40$} & \textcolor{blue}{\scriptsize{}$5727\pm\phantom{0}26$} & \textcolor{blue}{\scriptsize{}$5629$} & \textcolor{blue}{\scriptsize{}$5631$} & \multirow{2}{*}{\textcolor{blue}{\scriptsize{}$5637$}} & \textcolor{blue}{\scriptsize{}$5878\pm\phantom{0}97$} & \textcolor{blue}{\scriptsize{}$5803\pm\phantom{0}62$} & \textcolor{blue}{\scriptsize{}$5571$} & \textcolor{blue}{\scriptsize{}$5577$} & \multirow{2}{*}{\textcolor{blue}{\scriptsize{}$5599$}}\tabularnewline
 & \emph{\small{}B} & \textcolor{blue}{\scriptsize{}$5715\pm\phantom{0}28$} & \textcolor{blue}{\scriptsize{}$5699\pm\phantom{0}27$} & \textcolor{blue}{\scriptsize{}$5628$} & \textcolor{blue}{\scriptsize{}$5630$} &  & \textcolor{blue}{\scriptsize{}$5678\pm\phantom{0}88$} & \textcolor{blue}{\scriptsize{}$5603\pm\phantom{0}78$} & \textcolor{blue}{\scriptsize{}$5570$} & \textcolor{blue}{\scriptsize{}$5572$} & \tabularnewline
 & \emph{\small{}C} & \textcolor{blue}{\scriptsize{}$5614\pm\phantom{0}49$} & \textcolor{blue}{\scriptsize{}$5668\pm\phantom{0}66$} & \textcolor{blue}{\scriptsize{}$5628$} & \textcolor{blue}{\scriptsize{}$5629$} & \multirow{2}{*}{\textcolor{blue}{\scriptsize{}$\pm26$}} & \textcolor{blue}{\scriptsize{}$5674\pm122$} & \textcolor{blue}{\scriptsize{}$5635\pm\phantom{0}78$} & \textcolor{blue}{\scriptsize{}$5570$} & \textcolor{blue}{\scriptsize{}$5570$} & \multirow{2}{*}{\textcolor{blue}{\scriptsize{}$\pm25$}}\tabularnewline
\hline 
\end{tabular}
\par\end{centering}{\small \par}

\caption{\label{tab:Test3-4}Test 6. Delta calculation for the Dynamic Heston
case. All results must be multiplied by $10^{-4}$. The parameters
used for this test are available in Table \ref{tab:BS-CF-para}, \ref{tab:alpha-4}
and in Table \ref{tab:mp2}.}
\end{table}

\clearpage

\subsection{\label{5.3}Static Withdrawal and Optimal Surrender for GMWB-YD}

In the Static Withdrawal case we suppose the PH to withdrawal exactly
at the guaranteed rate, while in Optimal surrender case, the PH can
stop the contract at each event time.

The Tests 7 and 8 are inspired by \cite{YD}: in their article, Yang
and Dai price a GMWB contract both in Static and Dynamic (optimal
surrender) framework, under the Black Scholes model. First we priced
their products for different maturities and withdrawal rates, in Black
and Scholes model to get a reference price in this model and to compare
our results with the author's ones. We used a standard Monte Carlo
method and a standard PDE method for the Black Scholes model. Then,
we add stochastic volatility and stochastic interest rate. Model parameters
are available in Table \ref{tab:BS-YD}, and the values of $\alpha_{g}$
that we got are given in Table \ref{tab:BS-YD-alpha}.

We dealt with four numerical cases: deferred or not and Static behavior
or Surrendering.

We note that using different methods (a simple Monte Carlo approach,
and a PDE method for the Black-Scholes model) we didn't obtain the
same results of Yang and Dai in the case $\left(T_{1},T_{2}\right)=\left(10,25\right)$.
Probably we misunderstood some contract specifications about the deferred
case. We priced those products both using similarity reduction (see
Section \ref{sr}) and without, obtaining the same results. We would
remark that Yang and Dai didn't use this technique for their product.

\begin{table}
\begin{centering}
\begin{tabular}{|lr|r|rr|}
{\small{}Contract times $\left(T_{1},T_{2}\right)$} & {\small{}$\left(0,25\right)$ or $\left(10,25\right)$} &  & {\small{}GMW $G$} & $\frac{\max\left[P,W_{T_{1}}\right]}{T_{2}-T_{1}}$\tabularnewline
{\small{}Withdrawal Frequency $WF$} & {\small{}1 Year} &  & $m$ & {\small{}$1.0$}\tabularnewline
{\small{}Initial account value $A_{0}$} & {\small{}$100.0$} &  & {\small{}$S_{0}$} & {\small{}$100.0$}\tabularnewline
{\small{}Initial Premium} & {\small{}$100.0$} &  & {\small{}$r$} & {\small{}$0.0325$}\tabularnewline
{\small{}Withdrawal penalty $\kappa$} & {\small{}$0.10$} &  & {\small{}$\sigma$} & {\small{}$0.30$}\tabularnewline
PH's behavior & Static or Surrendering &  & Mortality & OFF\tabularnewline
\end{tabular}
\par\end{centering}

\caption{\label{tab:BS-YD}Parameters used by Yang and Dai in \cite{YD}.}

\vspace{1cm}

\begin{centering}
\begin{tabular}{c|ccc|ccc|}
\multirow{2}{*}{{\small{}$\left(T_{1},T_{2}\right)$}} & \multicolumn{3}{c|}{Static} & \multicolumn{3}{c|}{surrendering}\tabularnewline
 & {\small{}PDE} & {\small{}MC} & {\small{}YD} & {\small{}PDE} & {\small{}MC} & {\small{}YD}\tabularnewline
\hline 
\emph{\small{}$\left(0,25\right)$} & {\footnotesize{}$102.02$} & {\footnotesize{}$101.95\pm0.21$} & {\footnotesize{}$102$} & {\footnotesize{}$158.28$} & {\footnotesize{}$157.33\pm0.41$} & {\footnotesize{}$158$}\tabularnewline
\emph{\small{}$\left(10,25\right)$} & {\footnotesize{}$254.01$} & {\footnotesize{}$253.99\pm0.16$} & \textbf{\textcolor{red}{\footnotesize{}$170$}} & {\footnotesize{}$305.35$} & {\footnotesize{}$305.26\pm0.50$} & \textbf{\textcolor{red}{\footnotesize{}$248$}}\tabularnewline
\end{tabular}
\par\end{centering}

\caption{\label{tab:BS-YD-alpha}Fair bp values of $\alpha_{g}$ in Black Scholes
model, for GMWB-YD with the same parameters as in \cite{CF}.}
\end{table}

\subsubsection{Test 7: GMWB-YD in the Black-Scholes Hull-White Model}

In the conclusion of their paper \cite{YD}, Yang and Dai proposed
themselves to evaluate their contract including stochastic interest
rate. That's what we do in our paper, and in Test 7 we present some
numerical results about GMWB-YD pricing. Contract specifications are
shown in Table \ref{tab:BS-YD}, model parameters in Table \ref{tab:mp3}
and the fair values of $\alpha_{g}$ in Table \ref{tab:Test7}.

All four numerical methods behaved well in the Static case, but PDE
methods outperformed the others. Thing are different in the surrendering
case: the Longstaff Schwartz method showed its limits: in the BS HW
model the underlying and thus the account value can diffuse so much
in $25$ years and the regression over such a wide set of values is
stiff. PDE methods proved to be reliable and stable, especially in
case $\left(T_{1},T_{2}\right)=\left(0,25\right)$ where $25$ regressions
are required.

\subsubsection{Test 8: GMWB-YD in the Heston Model}

After pricing the GMWB-CF product in the BS and BS HW model, then
we did it in the Heston model. Contract specifications are shown in
Table \ref{tab:BS-YD}, model parameters in Table \ref{tab:mp4} and
the fair values of $\alpha_{g}$ in Table \ref{tab:Test8}. 

Like the previous test, all four numerical methods behaved well in
the Static case; HPDE and APDE outperformed the others and proved
to be equivalent in that framework. In this test, numerical results
of MC methods for the surrendering case are good: probably, the least
square regression is easier in the Heston case. Moreover, results
in the $\left(T_{1},T_{2}\right)=\left(10,25\right)$ case are very
good: in this case, the Longstaff-Schwartz algorithm requires only
$15$ numerical regressions and we can simulate more scenarios than
in the other case.

\clearpage

\begin{table}
\begin{centering}
{\small{}}%
\begin{tabular}{c|ccccc|c}
\hline 
{\small{}$S_{0}$} & {\small{}$r$} & {\small{}$curve$} & {\small{}$k$} & {\small{}$\omega$} & {\small{}$\rho$} & {\small{}$\sigma$}\tabularnewline
{\small{}$100$} & {\small{}$0.0325$} & {\small{}$flat$} & {\small{}$1.0$} & {\small{}$0.2$} & {\small{}$-0.5$} & {\small{}$0.30$}\tabularnewline
\hline 
\end{tabular}
\par\end{centering}{\small \par}

\caption{\label{tab:mp3}The model parameters about Test 7.}

\vspace{1cm}

\begin{centering}
\begin{tabular}{c||r}
\multicolumn{2}{c}{{\small{}}%
\begin{tabular}{|c|ccccc|ccccc|}
\cline{2-11} 
\multicolumn{1}{c|}{} & \multicolumn{5}{c|}{\emph{\small{}Static}} & \multicolumn{5}{c|}{\emph{\small{}surrendering}}\tabularnewline
\cline{2-11} 
\multicolumn{1}{c|}{} & {\footnotesize{}HMC} & {\footnotesize{}SMC} & {\footnotesize{}HPDE} & {\footnotesize{}APDE} & {\footnotesize{}BM} & {\footnotesize{}HMC} & {\footnotesize{}SMC} & {\footnotesize{}HPDE} & {\footnotesize{}APDE} & {\footnotesize{}BM}\tabularnewline
\cline{2-11} 
\multicolumn{11}{c}{}\tabularnewline
\multicolumn{11}{c}{{\small{}$\left(T_{1},T_{2}\right)=\left(0,25\right)$}}\tabularnewline
\hline 
\multirow{1}{*}{\emph{\small{}A}} & \textcolor{blue}{\scriptsize{}$83.11\pm3.55$} & \textcolor{blue}{\scriptsize{}$81.30\pm2.70$} & \textcolor{blue}{\scriptsize{}$80.79$} & \textcolor{blue}{\scriptsize{}$80.62$} & \multirow{2}{*}{\textcolor{blue}{\scriptsize{}$80.65$}} & \textcolor{blue}{\scriptsize{}$94.94\pm5.28$} & \textcolor{blue}{\scriptsize{}$89.98\pm4.34$} & \textcolor{blue}{\scriptsize{}$96.04$} & \textcolor{blue}{\scriptsize{}$95.98$} & \multirow{2}{*}{\textcolor{blue}{\scriptsize{}$92.95$}}\tabularnewline
\multirow{1}{*}{\emph{\small{}B}} & \textcolor{blue}{\scriptsize{}$83.06\pm2.48$} & \textcolor{blue}{\scriptsize{}$80.05\pm1.28$} & \textcolor{blue}{\scriptsize{}$80.71$} & \textcolor{blue}{\scriptsize{}$80.71$} &  & \textcolor{blue}{\scriptsize{}$89.12\pm2.20$} & \textcolor{blue}{\scriptsize{}$91.75\pm1.84$} & \textcolor{blue}{\scriptsize{}$95.50$} & \textcolor{blue}{\scriptsize{}$96.04$} & \tabularnewline
\multirow{1}{*}{\emph{\small{}C}} & \textcolor{blue}{\scriptsize{}$82.49\pm1.69$} & \textcolor{blue}{\scriptsize{}$80.62\pm0.57$} & \textcolor{blue}{\scriptsize{}$80.71$} & \textcolor{blue}{\scriptsize{}$80.72$} & \multirow{2}{*}{\textcolor{blue}{\scriptsize{}$\pm0.20$}} & \textcolor{blue}{\scriptsize{}$89.55\pm1.45$} & \textcolor{blue}{\scriptsize{}$91.79\pm1.34$} & \textcolor{blue}{\scriptsize{}$95.52$} & \textcolor{blue}{\scriptsize{}$96.08$} & \multirow{2}{*}{\textcolor{blue}{\scriptsize{}$\pm0.78$}}\tabularnewline
\emph{\small{}D} & \textcolor{blue}{\scriptsize{}$81.48\pm0.75$} & \textcolor{blue}{\scriptsize{}$80.80\pm0.28$} & \textcolor{blue}{\scriptsize{}$80.70$} & \textcolor{blue}{\scriptsize{}$80.72$} &  & \textcolor{blue}{\scriptsize{}$89.60\pm1.10$} & \textcolor{blue}{\scriptsize{}$90.22\pm1.11$} & \textcolor{blue}{\scriptsize{}$95.53$} & \textcolor{blue}{\scriptsize{}$96.09$} & \tabularnewline
\hline 
\multicolumn{11}{c}{}\tabularnewline
\multicolumn{11}{c}{{\small{}$\left(T_{1},T_{2}\right)=\left(10,25\right)$}}\tabularnewline
\hline 
\emph{\small{}A} & \textcolor{blue}{\scriptsize{}$213.24\pm3.05$} & \textcolor{blue}{\scriptsize{}$210.58\pm2.28$} & \textcolor{blue}{\scriptsize{}$210.40$} & \textcolor{blue}{\scriptsize{}$210.91$} & \multirow{2}{*}{\textcolor{blue}{\scriptsize{}$210.76$}} & \textcolor{blue}{\scriptsize{}$242.15\pm6.44$} & \textcolor{blue}{\scriptsize{}$233.16\pm5.58$} & \textcolor{blue}{\scriptsize{}$242.38$} & \textcolor{blue}{\scriptsize{}$242.86$} & \multirow{2}{*}{\textcolor{blue}{\scriptsize{}$241.41$}}\tabularnewline
\emph{\small{}B} & \textcolor{blue}{\scriptsize{}$212.68\pm2.13$} & \textcolor{blue}{\scriptsize{}$210.47\pm1.11$} & \textcolor{blue}{\scriptsize{}$210.67$} & \textcolor{blue}{\scriptsize{}$210.99$} &  & \textcolor{blue}{\scriptsize{}$244.06\pm3.38$} & \textcolor{blue}{\scriptsize{}$239.25\pm2.86$} & \textcolor{blue}{\scriptsize{}$242.83$} & \textcolor{blue}{\scriptsize{}$243.12$} & \tabularnewline
\emph{\small{}C} & \textcolor{blue}{\scriptsize{}$212.45\pm1.44$} & \textcolor{blue}{\scriptsize{}$210.72\pm0.49$} & \textcolor{blue}{\scriptsize{}$210.74$} & \textcolor{blue}{\scriptsize{}$210.89$} & \multirow{2}{*}{\textcolor{blue}{\scriptsize{}$\pm0.17$}} & \textcolor{blue}{\scriptsize{}$238.66\pm2.02$} & \textcolor{blue}{\scriptsize{}$239.37\pm1.67$} & \textcolor{blue}{\scriptsize{}$242.94$} & \textcolor{blue}{\scriptsize{}$243.07$} & \multirow{2}{*}{\textcolor{blue}{\scriptsize{}$\pm0.93$}}\tabularnewline
\emph{\small{}D} & \textcolor{blue}{\scriptsize{}$211.49\pm0.66$} & \textcolor{blue}{\scriptsize{}$210.73\pm0.25$} & \textcolor{blue}{\scriptsize{}$210.75$} & \textcolor{blue}{\scriptsize{}$210.84$} &  & \textcolor{blue}{\scriptsize{}$241.61\pm1.29$} & \textcolor{blue}{\scriptsize{}$239.75\pm1.27$} & \textcolor{blue}{\scriptsize{}$242.97$} & \textcolor{blue}{\scriptsize{}$243.04$} & \tabularnewline
\hline 
\end{tabular}}\tabularnewline
\end{tabular}
\par\end{centering}

\caption{\label{tab:Test7}Test 7. The fair fee $\alpha_{g}$ in bps for the
BS HW model, with Static withdrawal or Surrendering option. The parameters
used for this test are available in Table \ref{tab:BS-YD} and in
Table \ref{tab:mp3}.}
\end{table}

\begin{table}
\begin{centering}
{\small{}}%
\begin{tabular}{c|ccccc|c}
\hline 
{\small{}$S_{0}$} & {\small{}$v_{0}$} & {\small{}$\theta$} & {\small{}$k$} & {\small{}$\omega$} & {\small{}$\rho$} & {\small{}$r$}\tabularnewline
{\small{}$100$} & {\small{}$0.30^{2}$} & {\small{}$0.30^{2}$} & {\small{}$1.0$} & {\small{}$0.2$} & {\small{}$-0.5$} & {\small{}$0.0325$}\tabularnewline
\hline 
\end{tabular}
\par\end{centering}{\small \par}

\caption{\label{tab:mp4}The model parameters about Test 8.}

\vspace{1cm}

\begin{centering}
\begin{tabular}{c||r}
\multicolumn{2}{c}{{\small{}}%
\begin{tabular}{|c|ccccc|ccccc|}
\cline{2-11} 
\multicolumn{1}{c|}{} & \multicolumn{5}{c|}{\emph{\small{}Static}} & \multicolumn{5}{c|}{\emph{\small{}surrendering}}\tabularnewline
\cline{2-11} 
\multicolumn{1}{c|}{} & {\footnotesize{}HMC} & {\footnotesize{}SMC} & {\footnotesize{}HPDE} & {\footnotesize{}APDE} & {\footnotesize{}BM} & {\footnotesize{}HMC} & {\footnotesize{}SMC} & {\footnotesize{}HPDE} & {\footnotesize{}APDE} & {\footnotesize{}BM}\tabularnewline
\cline{2-11} 
\multicolumn{11}{c}{}\tabularnewline
\multicolumn{11}{c}{{\small{}$\left(T_{1},T_{2}\right)=\left(0,25\right)$}}\tabularnewline
\hline 
\multirow{1}{*}{\emph{\small{}A}} & \textcolor{blue}{\scriptsize{}$104.19\pm3.43$} & \textcolor{blue}{\scriptsize{}$104.49\pm3.64$} & \textcolor{blue}{\scriptsize{}$101.17$} & \textcolor{blue}{\scriptsize{}$101.10$} & \multirow{2}{*}{\textcolor{blue}{\scriptsize{}$100.71$}} & \textcolor{blue}{\scriptsize{}$142.75\pm4.67$} & \textcolor{blue}{\scriptsize{}$140.41\pm4.60$} & \textcolor{blue}{\scriptsize{}$145.58$} & \textcolor{blue}{\scriptsize{}$145.86$} & \multirow{2}{*}{\textcolor{blue}{\scriptsize{}$143.71$}}\tabularnewline
\multirow{1}{*}{\emph{\small{}B}} & \textcolor{blue}{\scriptsize{}$101.04\pm2.36$} & \textcolor{blue}{\scriptsize{}$102.43\pm2.62$} & \textcolor{blue}{\scriptsize{}$101.07$} & \textcolor{blue}{\scriptsize{}$101.07$} &  & \textcolor{blue}{\scriptsize{}$139.92\pm2.97$} & \textcolor{blue}{\scriptsize{}$138.92\pm2.82$} & \textcolor{blue}{\scriptsize{}$145.48$} & \textcolor{blue}{\scriptsize{}$145.80$} & \tabularnewline
\multirow{1}{*}{\emph{\small{}C}} & \textcolor{blue}{\scriptsize{}$101.49\pm1.30$} & \textcolor{blue}{\scriptsize{}$102.19\pm1.42$} & \textcolor{blue}{\scriptsize{}$101.07$} & \textcolor{blue}{\scriptsize{}$101.08$} & \multirow{2}{*}{\textcolor{blue}{\scriptsize{}$\pm0.52$}} & \textcolor{blue}{\scriptsize{}$141.57\pm1.55$} & \textcolor{blue}{\scriptsize{}$140.22\pm1.61$} & \textcolor{blue}{\scriptsize{}$145.61$} & \textcolor{blue}{\scriptsize{}$145.78$} & \multirow{2}{*}{\textcolor{blue}{\scriptsize{}$\pm0.57$}}\tabularnewline
\emph{\small{}D} & \textcolor{blue}{\scriptsize{}$101.45\pm0.75$} & \textcolor{blue}{\scriptsize{}$101.05\pm0.80$} & \textcolor{blue}{\scriptsize{}$101.07$} & \textcolor{blue}{\scriptsize{}$101.08$} &  & \textcolor{blue}{\scriptsize{}$142.04\pm1.05$} & \textcolor{blue}{\scriptsize{}$142.41\pm1.08$} & \textcolor{blue}{\scriptsize{}$145.62$} & \textcolor{blue}{\scriptsize{}$145.77$} & \tabularnewline
\hline 
\multicolumn{11}{c}{}\tabularnewline
\multicolumn{11}{c}{{\small{}$\left(T_{1},T_{2}\right)=\left(10,25\right)$}}\tabularnewline
\hline 
\emph{\small{}A} & \textcolor{blue}{\scriptsize{}$246.57\pm2.70$} & \textcolor{blue}{\scriptsize{}$248.46\pm2.90$} & \textcolor{blue}{\scriptsize{}$244.67$} & \textcolor{blue}{\scriptsize{}$244.45$} & \multirow{2}{*}{\textcolor{blue}{\scriptsize{}$244.52$}} & \textcolor{blue}{\scriptsize{}$280.93\pm5.33$} & \textcolor{blue}{\scriptsize{}$286.79\pm5.31$} & \textcolor{blue}{\scriptsize{}$286.20$} & \textcolor{blue}{\scriptsize{}$286.11$} & \multirow{2}{*}{\textcolor{blue}{\scriptsize{}$286.39$}}\tabularnewline
\emph{\small{}B} & \textcolor{blue}{\scriptsize{}$245.51\pm1.90$} & \textcolor{blue}{\scriptsize{}$248.04\pm2.11$} & \textcolor{blue}{\scriptsize{}$244.76$} & \textcolor{blue}{\scriptsize{}$244.68$} &  & \textcolor{blue}{\scriptsize{}$285.72\pm2.73$} & \textcolor{blue}{\scriptsize{}$286.46\pm2.88$} & \textcolor{blue}{\scriptsize{}$286.46$} & \textcolor{blue}{\scriptsize{}$286.42$} & \tabularnewline
\emph{\small{}C} & \textcolor{blue}{\scriptsize{}$245.31\pm1.03$} & \textcolor{blue}{\scriptsize{}$245.75\pm1.14$} & \textcolor{blue}{\scriptsize{}$244.80$} & \textcolor{blue}{\scriptsize{}$244.78$} & \multirow{2}{*}{\textcolor{blue}{\scriptsize{}$\pm0.41$}} & \textcolor{blue}{\scriptsize{}$286.67\pm1.70$} & \textcolor{blue}{\scriptsize{}$285.03\pm1.72$} & \textcolor{blue}{\scriptsize{}$286.56$} & \textcolor{blue}{\scriptsize{}$286.52$} & \multirow{2}{*}{\textcolor{blue}{\scriptsize{}$\pm0.65$}}\tabularnewline
\emph{\small{}D} & \textcolor{blue}{\scriptsize{}$245.42\pm0.60$} & \textcolor{blue}{\scriptsize{}$245.18\pm0.65$} & \textcolor{blue}{\scriptsize{}$244.81$} & \textcolor{blue}{\scriptsize{}$244.80$} &  & \textcolor{blue}{\scriptsize{}$286.54\pm1.01$} & \textcolor{blue}{\scriptsize{}$286.41\pm1.01$} & \textcolor{blue}{\scriptsize{}$286.57$} & \textcolor{blue}{\scriptsize{}$286.60$} & \tabularnewline
\hline 
\end{tabular}}\tabularnewline
\end{tabular}
\par\end{centering}

\caption{\label{tab:Test8}Test 8. The fair fee $\alpha_{g}$ in bps for the
Heston model, with Static withdrawal or Surrendering option. The parameters
used for this test are available in Table \ref{tab:BS-YD} and in
Table \ref{tab:mp4}.}
\end{table}

\clearpage

\section{\label{6}Conclusions}

In this article we have developed four numerical methods to price
two versions of GMWB contracts under different conditions. Regarding
the stochastic model, both stochastic interest rate and stochastic
volatility effects have been considered. Regarding the policy holder's
behavior, both static and dynamic strategy have been considered.

Since GMWB variable annuities are such a long maturity products, the
effects of stochastic interest rate and stochastic volatility cannot
be overlook. In particular, the impact of stochastic interest rate
seems to be more relevant. 

All four methods gave compatible results both for pricing and delta
calculation. The fair hedging fee (i.e. the cost of maintaining the
replicating portfolio) is determined using a sequence of parameters
refinements. The PDE methods proved to be not very expensive, while
MC methods proved to be more expensive. The Hybrid PDE seemed to be
the more performing than the others for its convergence speed and
stability of results. Also ADI PDE behaved very well but the implementation
was a little harder than Hybrid PDE one; moreover the choice of the
good parameters for ADI PDE was a source of troubles. In the BS HW
model case, Standard MC, thanks to its exact simulation, outperformed
the hybrid method while, in the Heston model case, the MC methods
proved to be roughly equivalent, even if the Hybrid MC was easier
to be implemented.

As we said before, PDE methods proved to be much more efficient than
MC methods, especially in Dynamic case where it's much more simple
to implement the optimal withdrawal choice. In the GMWB-YD case, similarity
reduction reduces the dimension of the problem to 2 and therefore
PDE methods perform very well. In the GMWB-CF case similarity reduction
cannot be applied and therefore pricing is an harder task, especially
in the case of six-monthly withdrawal and $20$ years maturity. Anyway,
we have to remark that MC methods offer a confidence interval for
results, they are useful in risk measures calculation (for example
VAR or ES), and they are preferred by insurance companies because
of their attachment to the idea of scenario.

The use of special numerical techniques (splines, improved LS convergence)
allowed to ensure the convergence containing the computational time.

A future development that could be treated is to combine stochastic
interest rate and stochastic volatility: the combined model could
be an element of greater realism.

We conclude by pointing out that our methods are quite flexible in
that they can accommodate a wide variety of policy holder withdrawal
strategies such as ones derived from utility-based models.

\end{document}